\documentclass[prd,preprintnumbers,superscriptaddress,amsmath,amssymb,nofootinbib,twocolumn,floatfix]{revtex4-2}

\usepackage{graphicx}
\usepackage{dcolumn}
\usepackage{bm}
\usepackage{orcidlink}

\usepackage{ulem}

\begin{document}

\preprint{APS/123-QED}
\preprint{KCL-PH-TH-2026-18}

\title{Parameter Estimation on LIGO-Virgo-KAGRA O4a Binary Merger Triggers with  Sub-solar Mass Components}
\author{Nelson Christensen,\orcidlink{0000-0002-6870-4202}}
 \email{nelson.christensen@oca.eu}
 \affiliation{
 Université Côte d’Azur, Observatoire de la Côte d’Azur, CNRS, Laboratoire Artemis, 06300 Nice, France}
 
 \author{Neil Cornish,\orcidlink{0000-0002-7435-0869}}
 \email{ncornish@montana.edu}
 \affiliation{eXtreme Gravity Institute, Department of Physics,
Montana State University, Bozeman, MT 59717, USA}

\author{Florian K\"{u}hnel,\orcidlink{0000-0002-1528-1920}}
 \email{fkuehnel@mpp.mpg.de}
 \affiliation{
Max Planck Institute for Physics, Boltzmannstr. 8, 85748 Garching, Germany}
 
\author{Mairi Sakellariadou,\orcidlink{0000-0002-2715-1517}}
\email{mairi.sakellariadou@kcl.ac.uk}
\affiliation{Theoretical Particle Physics and Cosmology Group, Physics Department, King’s College London,
University of London, Strand, London WC2R 2LS, United Kingdom\\ }
\affiliation{
 Université Côte d’Azur, Observatoire de la Côte d’Azur, CNRS, Laboratoire Artemis, 06300 Nice, France}

\author{Andres Santiago Villares Guanga,\orcidlink{0000-0003-2864-3967}} 
 \email{andres.villaresguanga@oca.eu}
  \affiliation{
 Université Côte d’Azur, Observatoire de la Côte d’Azur, CNRS, Laboratoire Artemis, 06300 Nice, France}

\date{\today}

\begin{abstract}
The LIGO-Virgo-KAGRA collaboration has reported the results for searches for sub-solar mass components in compact binary coalescence during observing run O4a. No detection candidates were identified, but the most significant seven triggers were reported. We present the results of Bayesian parameter inference on these triggers. Five of the triggers show agreement between the Bayesian parameter estimation and the search pipeline trigger. Our results show that three of the triggers may contain a possible sub-solar mass component. Parameter estimation indicates that the other two events, if real, would be neutron star - black hole binaries. The remaining two triggers do not provide informative parameter estimation. We also study three O4a compact binary coalescence triggers, and one O3 trigger, that have been noted by three other groups, and our parameter estimation indicates that three of these may contain a sub-solar mass component. We study the data quality associated with these triggers. Finally, we discuss the challenges for parameter estimation on compact binary coalescence events containing a sub-solar mass component: long signal duration, possible small chirp masses, possible small mass ratios, and data quality issues over potentially hundreds of seconds of data.
\end{abstract}

\maketitle



\section{Introduction} \label{sec:intro}
The LIGO-Virgo-KAGRA collaboration (LVK)~\cite{LIGOScientific:2014pky,VIRGO:2014yos,KAGRA:2020tym} has recently reported 390 significant detections of gravitational waves from compact binary mergers~\cite{LIGOScientific:2026wfs}. This covers the data from observing runs O1, O2, O3, O4a and O4b~\cite{LIGOScientific:2026sit}. Of these two were from binary neutron star mergers~\cite{LIGOScientific:2017vwq,LIGOScientific:2020aai}, while six were likely from neutron star - black hole mergers~\cite{LIGOScientific:2021qlt,LIGOScientific:2024elc,LIGOScientific:2020zkf}. The remaining events were produced by binary black hole mergers. 

The LVK also conducts searches for compact binary mergers where at least one of the components would have a mass smaller than $1 \, \textup{M}_\odot$~\cite{LIGOScientific:2005fbz,LIGOScientific:2018glc,LIGOScientific:2019kan,LIGOScientific:2021job,LVK:2022ydq}. The LVK has also searched the O4a data for such events~\cite{LIGOScientific:2026XXX}. No detections were reported, but the seven most significant triggers were presented, along with the component masses and spins corresponding to the triggers.

The observation of a sub-solar mass compact object would be of tremendous importance for physics and cosmology. It is difficult to explain the formation of a sub-solar mass black hole by stellar processes. Similarly, a sub-solar mass neutron star would need a non-standard equation of state~\cite{Carr:2023tpt}. Hence, the observation of such an object would most naturally point to a {\it primordial black hole} (PBH)~\cite{Hawking:1971ei, Carr:1974nx}. PBHs are black holes formed in the early Universe, rather than as the endpoint of stellar evolution, and have been the subject of an extensive literature, especially in their role as potential dark matter candidates and as probes of early-Universe physics (see, for instance, Refs.~\cite{Carr:2016drx, Carr:2020xqk, Escriva:2022duf} for recent reviews). Their phenomenology is exceptionally broad, involving formation from enhanced primordial perturbations, phase transitions or other early-Universe mechanisms, and observational signatures ranging from gravitational lensing and dynamical effects to accretion, evaporation and gravitational waves.

At present, there are many intriguing hints for the existence of PBHs. These include microlensing events in several mass ranges, in particular quasar microlensing~\cite{2017ApJ...836L..18M}, including systems so misaligned with the lensing galaxy that stellar microlensing is strongly disfavoured, as well as the excess source-subtracted near-infrared cosmic infrared background anisotropies~\cite{Kashlinsky:2005di} (see also \cite{Cappelluti:2013bda, Kashlinsky:2016sdv, Hasinger:2020ptw}). 
Further support comes from high-redshift observations, especially the unexpectedly early appearance of massive galaxies and black holes in JWST data. Examples include the luminous JADES-GS-$z14$ galaxies at $z \simeq 14$ \cite{2024Natur.633..318C} and the over-massive active nucleus Abell2744-QSO1 at $z \simeq 7.04$, which contains a $\simeq 5 \times 10^{7}\,\textup{M}_{\odot}$ black hole, has an extreme black-hole-to-stellar mass ratio $M_{\rm BH} / M_{\star} > 2$, and appears to reside in a very low-metallicity, nearly pristine environment with $Z\lesssim 5 \times 10^{-3}\,Z_{\odot}$ \cite{2024Natur.628...57F, Juodzbalis:2025qso1, Maiolino:2025qso1}. Such systems are difficult to reconcile with ordinary stellar-seed growth within the short cosmic time available, whereas massive PBH seeds provide a natural and economical explanation~\cite{Zhang:2025oyl}.

PBHs can be produced over an enormous mass range, and their mass function is generically expected to be extended rather than monochromatic. One particularly natural and unified scenario is the thermal-history model of Ref.~\cite{Carr:2019kxo}, in which changes in the effective equation of state of the early Universe lower the sound speed and thereby enhance PBH formation at preferred horizon masses. In this picture, well-understood Standard Model epochs, including the electroweak transition, the quantum chromodynamics (QCD) crossover, the pion plateau and $e^+e^-$ annihilation, induce characteristic peaks in the PBH mass function around $10^{-6},1,\,30,$ and $10^{6}\,\textup{M}_{\odot}$, respectively. This multi-modal mass function is especially attractive because it can simultaneously address several apparently unrelated observational anomalies, including planetary-mass and stellar-mass microlensing events, mass-gap and high-mass gravitational-wave sources, early supermassive black hole seeds, and the rapid emergence of high-redshift structures \cite{Carr:2019kxo, Carr:2023tpt}. In particular, the low-mass side of the QCD peak naturally produces an appreciable population of sub-solar PBHs. A confirmed sub-solar compact object would require careful interpretation, since a PBH should be effectively point-like during the inspiral, whereas a sub-solar neutron star or other exotic compact object could in principle be distinguished through finite-size effects, most notably tidal deformability or tidal disruption.

In this paper we present the results of Bayesian parameter inference~\cite{Christensen:2022bxb} on the seven most significant triggers presented in the LVK search of O4a data~\cite{LIGOScientific:2026XXX}. We use the parameter inference method presented in~\cite{Cornish:2021wxy}. Of the seven triggers, three might possibly contain a sub-solar mass black hole. The estimated masses and spins are presented. We also investigate the LVK sub-threshold trigger GW231109\_235456~\cite{LIGOScientific:2025slb} (low-latency alert S231109ci), which was studied and reported to be a binary neutron star merger~\cite{Niu:2025nha}. In addition, another group searched the O4a data for compact binary coalescences where one component would be sub-solar mass~\cite{Kacanja:2026byy}. No significant events were found but the two more significant triggers were reported. Finally, there has been a parameter estimation study of an O3 compact binary coalescence event, SSM200308, where both of the compact objects might be sub-solar mass~\cite{Prunier:2023uoo}. Our parameter estimation results agree with this earlier analysis. We use these additional four events as an opportunity to test our parameter estimation methods.

The statistical significances for these triggers are given from the signal searches. For some of the events studied, the parameter estimation results support the possibility that the compact binary system contains a sub-solar mass object. We analyze the LIGO Hanford (H1) and LIGO Livingston (L1) data together. However, to give support for the parameter estimation conclusions, we also analyze the H1 and L1 data individually. When the parameter estimation results for H1 and L1 alone differ from that of the combined H1-L1 analysis, it argues against the event being real. We note that for the O3 trigger, SSM200308, we analyze H1, L1 and Virgo data. For O4a there are only H1 and L1 data.

In order to understand the ability to correctly estimate the parameters for a low mass compact binary system, we inject artificial signals into the H1 and L1 data. The results of this injection study are presented below.

Finally, we address the complexities of parameter estimation for compact binary coalescence events that contain a sub-solar mass compact object. These low mass events are in the sensitive band of the detectors for hundreds of seconds, requiring the analysis to use very long stretches of data. The stretches of data often contain non-Gaussian noise transients, or glitches, and may also exhibit longer term non-stationarity.  We present an analysis of the data quality for the triggers that we have studied. A very low chirp mass, or a small mass ratio, can be problematic for waveforms used in the parameter estimation; this is discussed. Other studies have also investigated the ability for the LVK to observe binary coalescence events that contain a sub-solar mass compact object~\cite{Wolfe:2023yuu,Newell:2026cma}.

The paper is organized as follows. The triggers studied are presented in Sec.~\ref{sec:candidates}. The parameter estimation methods are described in Sec.~\ref{sec:methods}. Our parameter estimation results are given in Sec.~\ref{sec:results}, including the results on single detector data in Sec.~\ref{subsec:single-det-results}. The data quality study is described in Sec.~\ref{sec:dq}. Our conclusions are presented in Sec.~\ref{sec:conclusions}. The studies pertaining to the triggers generated by non-LVK groups are presented in Appendix~\ref{App:others}. Supplemental figures from various studies presented here appear in Appendix~\ref{App:SuppFigs}. The parameter estimates that are given in this publication are the symmetric 90\% credible intervals.

\section{Candidate List} \label{sec:candidates}
The LVK searched the O4a data for binary mergers where one component might be a sub-solar mass object~\cite{LIGOScientific:2026XXX}. Three pipelines were used: GSTLAL~\cite{Hanna:2022zpk,Joshi:2025zdu,Joshi:2025nty}, MBTA~\cite{Aubin:2020goo,Allene:2025saz} and PYCBC~\cite{Davies:2020tsx,LIGOScientific:2025yae}. GSTLAL uses templates using IMRPhenomD waveforms~\cite{Khan:2015jqa,Husa:2015iqa}, while the analysis uses TaylorF2~\cite{Buonanno:2009zt,Boyle:2009dg} and SEOB-NRv4ROM~\cite{Buonanno:2009zt,PhysRevD.101.124040} approximants. 
Specifically, template bank generation and parameter-space placement in GSTLAL are performed using the IMRPhenomD waveform model. For the actual matched-filtering analysis, GSTLAL evaluates the data using TaylorF2 for low-mass systems, where redshifted chirp mass $(1+z)\mathcal{M} \le 1.73\,M_\odot$, and SEOBNRv4ROM for higher-mass templates to balance computational efficiency with waveform accuracy.
For MBTA, the template bank is made with TaylorF2~\cite{Buonanno:2009zt,Boyle:2009dg}, while the analysis is done with the SpinTaylorT4~\cite{Isoyama2020} approximant. PYCBC placed its templates using a generic algorithm, but then used TaylorF2~\cite{Buonanno:2009zt,Boyle:2009dg} and SEOB-NRv4ROM~\cite{Buonanno:2009zt,PhysRevD.101.124040}. The LVK's O4a sub-solar mass search targeted compact binary coalescences where the primary mass (in the detector frame) $m_{1}$ is between $0.2 \, \textup{M}_\odot$ and $10 \, \textup{M}_\odot$, and the secondary mass (detector frame) is between $0.1 \, \textup{M}_\odot$ and $1 \, \textup{M}_\odot$.

There was no definitive detection by the LVK of a compact binary coalescence containing a sub-solar mass black hole during O4a. However, the loudest seven triggers, with false alarm rates below 2 per year, were reported~\cite{LIGOScientific:2026XXX}. These are presented in Table~\ref{tab:table_triggers}.

Since the LVK search region explicitly includes secondary masses below $1\,M_\odot$, these triggers are not merely low-significance binary-merger candidates, but targeted probes of the mass range in which a black hole interpretation would be difficult to accommodate astrophysically and would naturally suggest a PBH origin.

\begin{table*}
\caption{\label{tab:table_triggers} The seven unique triggers from the LVK O4a sub-solar mass search~\cite{LIGOScientific:2026XXX}. These are the triggers with a false alarm rate less than two per year. The events are ordered by the false alarm rate (FAR).}
\begin{ruledtabular}
\begin{tabular}{ccccccc}
 Time&Primary mass $m_{1}$&Secondary mass $m_{2}$&Chirp mass&Detectors
&Network SNR&FAR \\ 
(UTC)& Detector Frame ($\textup{M}_\odot$)&Detector Frame ($\textup{M}_\odot$)&Detector Frame ($\textup{M}_\odot$) & & & yr$^{-1}$ \\ \hline
2023-05-29 18:15:00.75&6.49&0.98 &2.030&L&10.63&0.00052\footnote{All three pipelines observed this event, and we use the values from GSTLAL, which had the lowest false-alarm-rate.} \\
 2023-09-14 21:02:19.50&2.63&0.29&0.686&HL&9.44&0.87\\
 2023-05-28 11:07:32.09&6.49&0.98&2.030&HL&9.48&1.2\\
 2023-07-19 11:50:50.27&0.74&0.24&0.356&HL&9.47&1.3\\
 2023-10-14 08:15:06.33&2.28&0.22&0.550&HL&10.28&1.3\\
 2023-08-10 07:13:47.88&1.86&0.32&0.627&L&12.16&1.3\\
 2023-08-10 10:10:03.37&0.60&0.21&0.301&HL&9.43&1.4
\end{tabular}
\end{ruledtabular}
\end{table*}

As a means of verifying our method we also study the sub-threshold LVK trigger GW231109\_235456, a study of which was presented in~\cite{Niu:2025nha}. The source frame masses for the binary were estimated to be in the range 1.40 $\textup{M}_\odot$ to 2.24 $\textup{M}_\odot$ for the primary and 0.97 $\textup{M}_\odot$ to 1.49 $\textup{M}_\odot$ for the secondary. The detector frame chirp mass was estimated to be $m_c = 1.3063^{+0.0003}_{-0.0003} ~ \textup{M}_\odot$.

We study two triggers from the sub-solar mass compact binary search presented in~\cite{Kacanja:2026byy}. While no significant events where found, the top two detector candidates were reported, with detector frame chirp masses of $m_c = 0.32 ~ \textup{M}_\odot$ and $m_c = 0.30 ~ \textup{M}_\odot$.

Finally, we also analyze SSM200308, which was reported as an O3 GSTLAL trigger by the LVK~\cite{LVK:2022ydq}, and studied with parameter estimation in~\cite{Prunier:2023uoo}. The parameter estimation reported a detector frame chirp mass of $m_c = 0.3527^{+0.0003}_{-0.0001} ~ \textup{M}_\odot$, source frame component masses of $m_1 = 0.62^{+0.46}_{-0.20} ~ \textup{M}_\odot$ and $m_2 = 0.27^{+0.12}_{-0.10} ~ \textup{M}_\odot$, and effective spin of $\chi_{\rm eff} = 0.41^{+0.08}_{-0.04}$.

\section{Methods}\label{sec:methods}
We use the Bayesian parameter inference method, QuickCBC, of Cornish~\cite{Cornish:2021wxy}. The algorithm can de-noise and remove glitches from the data. The noise power spectral density is calculated on-source via an iterative method that involves the de-noising. A parallel tempered Markov chain Monte Carlo algorithm~\cite{PhysRevLett.57.2607} is used to perform the parameter estimation. The IMRPhenomD~\cite{Husa:2015iqa,Khan:2015jqa} model is used for the waveforms. Here we report the estimates of the source frame masses, $m_1$ and $m_2$, the detector frame chirp mass, $m_c$, the luminosity distance, $D_L$, the effective spin, $\chi_{\textup{eff}}$, the signal-to-noise ratio in each interferometer, and the network signal-to-noise ratio. We ignore possible tidal effects. 

We run QuickCBC with two different sets of priors depending on the expected total mass of the binary system. If the total mass of the system is expect to be less than $6 \, \textup{M}_\odot$, then the priors are: total mass $0.4 \, \textup{M}_\odot \le m_t \le 6.0 \, \textup{M}_\odot$; chirp mass $0.25 \, \textup{M}_\odot \le m_c \le 2.6 \, \textup{M}_\odot$ (but $m_c$ is at most 0.435 times $m_t$); the magnitude of the dimensionless spin of each mass $\chi_i \le 0.9$; 1 Mpc $\le D_L \le $ 1000 Mpc. When the total mass is likely to be above $6 \, \textup{M}_\odot$, then the priors are: total mass $1.0 \, \textup{M}_\odot \le m_t \le 500 \, \textup{M}_\odot$; chirp mass $1.0 \, \textup{M}_\odot \le m_c \le 200 \, \textup{M}_\odot $ (but $m_c$ is at most 0.435 times $m_t$); the magnitude of the dimensionless spin of each mass $\chi_i \le 0.95$ ; 1 Mpc $\le D_L \le $ 1000 Mpc. For both settings there are also the priors: $m_2/m_1 \ge 0.05$ and 1 s width for the merger time.

Different lengths of data are used for the analysis depending on the mass of the system. Lighter systems used up to 512 s of data, while heavier systems used 32 s. We note the amount of data used when presenting the parameter estimation results.

\section{Results}\label{sec:results}
The results of the parameter estimation are presented in Tables~\ref{tab:table_PE} and \ref{tab:mc_and_chi}.

\begin{table*}
\caption{\label{tab:table_PE} Parameter estimation results for the events listed in Table~\ref{tab:table_triggers}, for those events for which the parameter estimation indicated that they might be real gravitational-wave signals. The primary and secondary masses are for the source frame. The signal-to-noise ratio (SNR) is given for each detector individually (H for LIGO Hanford, and L for LIGO Livingston). The Network SNR is their quadrature sum. The length of data used is noted.}
\begin{ruledtabular}
\begin{tabular}{cccccccc}
 Time& $m_{1}$ ($\textup{M}_\odot$)& $m_{2}$ ($\textup{M}_\odot$)&$D_L$ (Mpc)&H SNR
&L SNR&Network SNR& Data length (s) \\  \hline
2023-05-29 18:15:00.75&$3.92^{+0.61}_{-0.56}$&$1.32^{+0.18}_{-0.14}$&$254.26^{+66.91}_{-88.39}$& &10.64&10.64&256 \\
 2023-09-14 21:02:19.50&$2.6164^{+0.1923}_{-0.1561}$&$0.2722^{+0.0130}_{-0.0137}$&$169.25^{+54.53}_{-60.56}$&6.58&6.16&9.01&256\\
 2023-05-28 11:07:32.09&$3.29^{+0.70}_{-0.62}$&$1.41^{+0.31}_{-0.20}$&$443.97^{+137.45}_{-135.25}$&6.18&4.98&7.94&32\\
 2023-07-19 11:50:50.27&$0.8185^{+0.063}_{-0.0581}$&$0.2185^{+0.0138}_{-0.0113}$&$98.18^{+23.88}_{-28.93}$&7.13&6.13&9.40&512\\
 2023-10-14 08:15:06.33&$2.17328^{+0.34568}_{-0.31304}$&$0.22673^{+0.02768}_{-0.02281}$&$58.97^{+37.33}_{-29.58}$&6.39&6.24&8.93&128
\end{tabular}
\end{ruledtabular}
\end{table*}

For the five triggers where we can estimate parameters there is a good agreement between the estimates of the detector frame chirp masses and the chirp masses from the triggers~\cite{LIGOScientific:2026XXX}. Three of the triggers, if real events, would correspond to a compact binary systems with a sub-solar mass component object, while the other two triggers would correspond to light neutron star - black hole binaries. 

\begin{figure}[tb]
        \centering
    	\includegraphics[width=\columnwidth]{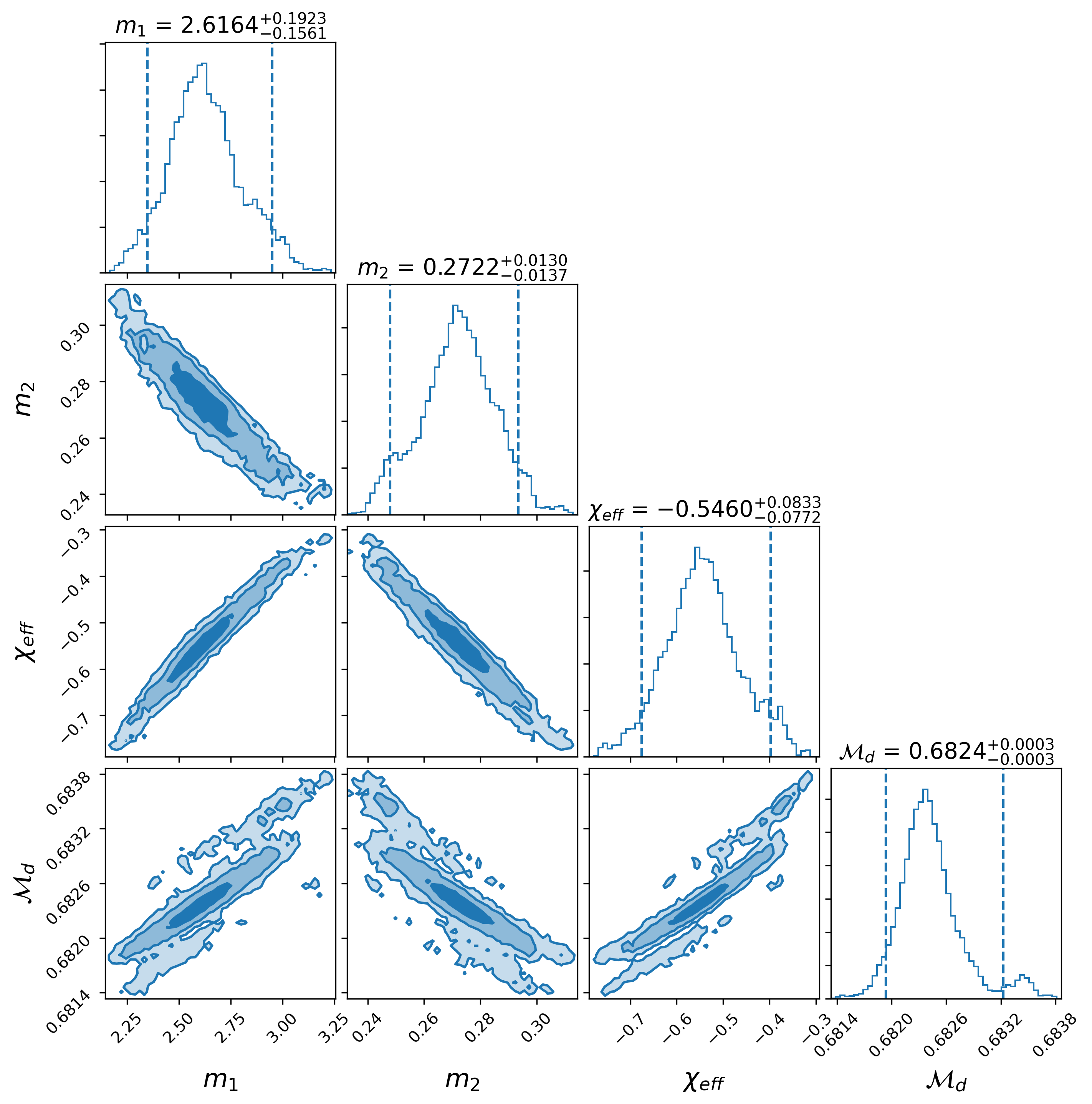}
    	\caption{A corner plot for the trigger at 2023-09-14 21:02:19.50 UTC showing the source frame component masses, $m_1$ and $m_2$, the effective spin $\chi_{\rm eff}$, and the detector frame chirp mass $m_c$.}
        \label{fig:1378760557_256_2}
    \end{figure}

The trigger at 2023-09-14 21:02:19.50 UTC is estimated to be a binary with one compact object at $m_2 \sim 0.27 ~ \textup{M}_\odot$. A corner plot showing the source from component masses, the detector frame chirp mass, and the effective spin are displayed in Fig.~\ref{fig:1378760557_256_2}. The mass estimates for this trigger are consistent with those from the signal search pipeline, and show a possible sub-solar mass object.

\begin{figure}[tb]
        \centering
    	\includegraphics[width=\columnwidth]{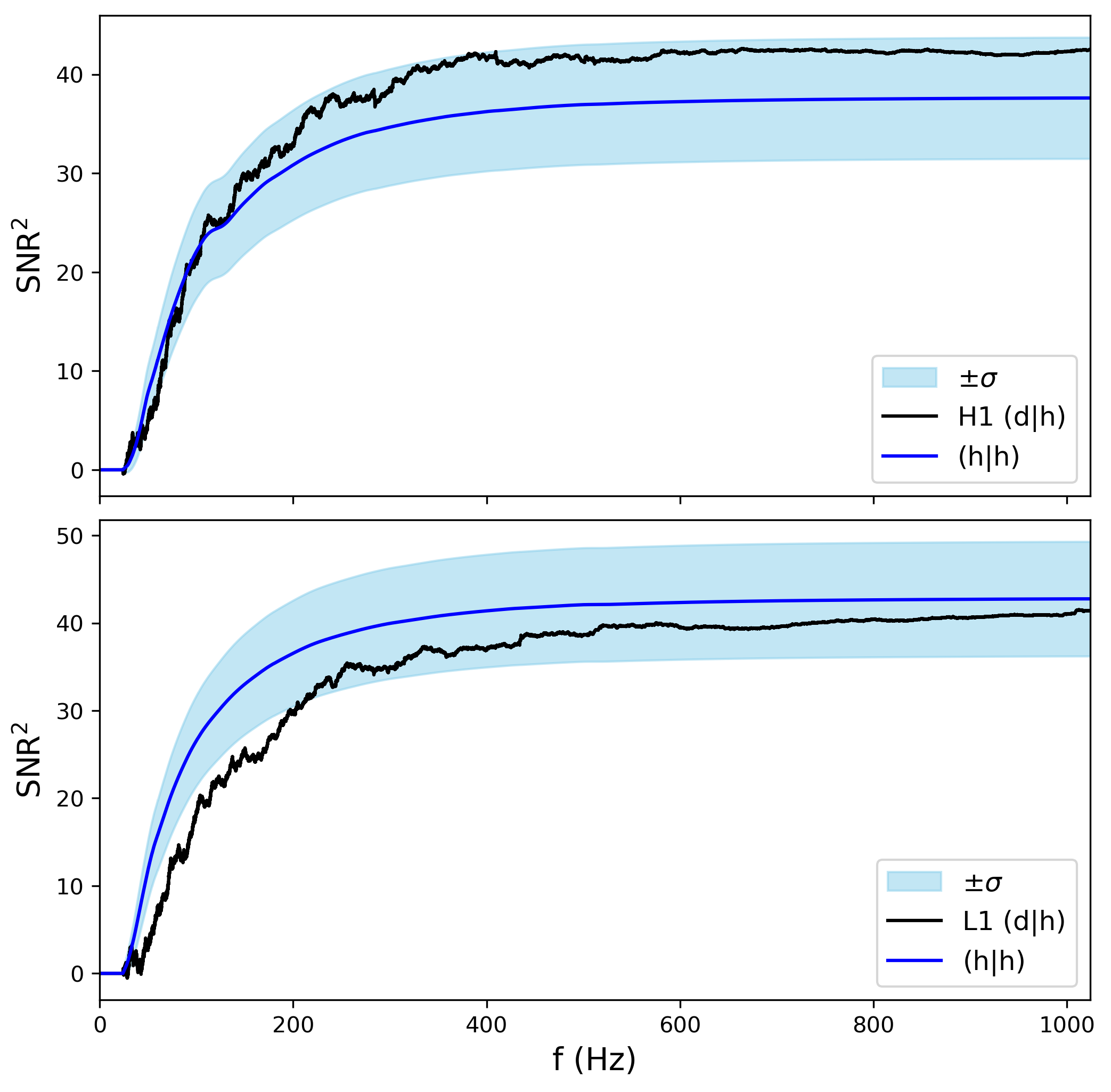}
    	\caption{The SNR$^2$ of the signal as a function of frequency for the  2023-09-14 21:02:19.50 UTC trigger. The black line is calculated from the data ($d \mid h$). The blue line corresponds to a draw from the likelihood ($h \mid h$). The blue region are the $\pm 1 \sigma$ errors on the SNR$^2$ estimation.}
        \label{fig:1378760557_SNR2_H1L1_together}
    \end{figure}

From the output of QuickCBC one can also calculate the SNR of a signal in the H1 and L1 data streams. One can also calculate the SNR$^2$ as a function of frequency. In Fig.~\ref{fig:1378760557_SNR2_H1L1_together} one can see SNR$^2(f)$, as well as the expected SNR$^2(f)$ from a draw from the likelihood in the QuickCBC analysis of the combined H1-L1 data. For this particular trigger the form of SNR$^2(f)$ matches reasonably well with the expected functional form for a compact binary coalescence with masses given in Table~\ref{tab:table_PE}.

The trigger at 2023-07-19 11:50:50.27 UTC is estimated to be a binary with compact objects of $m_1 \sim 0.82 ~ \textup{M}_\odot$ and $m_2 \sim 0.22 ~ \textup{M}_\odot$. A corner plot showing the source from component masses, the detector frame chirp mass, and the effective spin are displayed in Fig.~\ref{fig:1373802668_256_2} in the Appendix. The mass estimates for this trigger are consistent with those from the signal search pipeline, and show a possible sub-solar mass object. 
In Fig.~\ref{fig:1373802668_SNR2_H1L1_together} in the Appendix one can see that  the SNR$^2(f)$ accumulation similarly behaves as expected for a CBC signal with masses given in Table~\ref{tab:table_PE}

As described above, when we run the parameter estimation on combined H1-L1 data we use a prior on the chirp mass of $0.25 \, \textup{M}_\odot \le m_c \le 2.6 \, \textup{M}_\odot$. For the trigger 2023-10-14 08:15:06.33 UTC and using this chirp mass prior we were unable to generate good parameter estimation results. We do note that when the prior was constrained to $0.5 \, \textup{M}_\odot \le m_c \le 0.6 \, \textup{M}_\odot$ we could produce results. This trigger is estimated to be a binary with compact objects of $m_1 \sim 2.17 ~ \textup{M}_\odot$ and $m_2 \sim 0.23 ~ \textup{M}_\odot$. A corner plot showing the source from component masses, the detector frame chirp mass, and the effective spin are displayed in Fig.~\ref{fig:1381306524_256_2} in the Appendix. The mass and spin estimates for this trigger are consistent with those from the signal search pipeline, and show a possible sub-solar mass object. As seen in Fig.~\ref{fig:1381306524_SNR2_H1L1_together} in the Appendix, the SNR$^2(f)$ accumulation is consisten with the expected CBC signal with masses given in Table~\ref{tab:table_PE}.

If any of the three O4a triggers for which we find $m_{2} \lesssim 1\,\textup{M}_{\odot}$ were ultimately confirmed as genuine compact-binary coalescences, their interpretation would already be highly non-standard. Ordinary stellar evolution does not provide a natural channel for producing black holes below the Chandrasekhar scale, and core collapse supernova simulations set the theoretical minimum neutron star mass at $1.192\mathcal{M}_{\odot}$~\cite{müller2025minimumneutronstarmass}. Consequently, a sub solar neutron star would require an unusual formation history. Among the proposed non standard formation channels is the fragmentation of self-gravitating neutrino-cooled collapsar disk, which may produce neutron star binaries with sub solar masses ~\cite{chen2025gravitationalinstabilityfragmentationcollapsar}.


PBHs, on the other hand, can form below $\sim 1\,M_{\odot}$ without being subject to any stellar-evolution mass threshold. While candidates identified by a sub-solar matched filter search naturally reflect this selection effect, it is noteworthy that full Bayesian parameter estimation recovers well bounded posterior modes within the sub-solar regime, rather than scattering into uninformative or glitch-like distributions. From a PBH perspective, these events remain interesting probes of the sub-solar parameter space.

As mentioned in the Introduction, this mass range is also theoretically well motivated. In PBH scenarios in which the formation probability is enhanced around the QCD epoch, the softening of the equation of state can generate a broad mass function centred near the solar-mass scale, with an appreciable low-mass tail extending into the sub-solar regime. The candidate systems studied here therefore probe precisely the region in which a primordial interpretation would be natural, rather than merely representing a formal extension of compact-binary searches to low component masses.

There is some correspondence between the triggers' spin values and the parameter estimation results. For the trigger at 2023-09-14 21:02:19.50 the LVK reports dimensionless spin magnitudes of $\chi_1 = -0.72$ and $\chi_2 = -0.02$. Using 
\begin{equation}\label{eq:chi}
    \chi_{\rm eff}=\frac{m_1 \chi_1 + m_2 \chi_2}{m_1 + m_2}
\end{equation}
the effective spin is calculated to be $\chi_{\rm eff} = -0.65$, while parameter estimation gives $\chi_{\rm eff} = -0.546$. For the trigger at 2023-07-19 11:50:50.27 UTC the trigger gives $\chi_{\rm eff} = -0.157$, while parameter estimation gives $\chi_{\rm eff} = -0.09$. For the trigger at 2023-10-14 08:15:06.33 UTC the trigger gives $\chi_{\rm eff} = 0.83$, while parameter estimation gives $\chi_{\rm eff} = 0.80$.

\begin{table*}
\caption{\label{tab:mc_and_chi} Presented here are the comparisons of the spins and detector frame chirp masses as given by the triggers, given in Table~\ref{tab:table_triggers}, and the parameter estimation (PE).  The effective spin is estimated for the triggers from Eq.~\ref{eq:chi}.}
\begin{ruledtabular}
\begin{tabular}{ccccc}
 Time& $m_{c}$ ($\textup{M}_\odot$) trigger& $m_{c}$ ($\textup{M}_\odot$) PE&$\chi_{\textup{eff}}$ trigger&$\chi_{\textup{eff}}$ PE \\  \hline
2023-05-29 18:15:00.75&2.030&$2.027^{+0.001}_{-0.001}$&0.27&$-0.04^{+0.09}_{-0.09}$ \\
 2023-09-14 21:02:19.50&$0.686$&$0.6824^{+0.0003}_{-0.0003}$&-0.65&$-0.5460^{+0.0833}_{-0.0772}$ \\
 2023-05-28 11:07:32.09&2.030&$2.017^{+0.002}_{-0.002}$&0.73&$0.47^{+0.06}_{-0.05}$\\
 2023-07-19 11:50:50.27&0.356&$0.3609^{+0.00005}_{-0.00004}$& -0.157 &$-0.0906^{+0.0429}_{-0.0432}$\\
 2023-10-14 08:15:06.33&0.550&$0.55659^{+0.00040}_{-0.00036}$&0.8252&$0.80122^{+0.02896}_{-0.02118}$
\end{tabular}
\end{ruledtabular}
\end{table*}

\subsection{Analysis of single detector data}
\label{subsec:single-det-results}
As a data quality and consistency check, we analyzed the data from individual detectors to see if there was agreement with the combined analysis. In this way we compare the parameter estimation results on the combined H1-L1 data with the results from H1 and L1 individually. This is difficult in that the signal-to-noise ratio for individual detectors tends to be about $\sqrt{2}$ smaller than the network signal to noise ratio. However, we look for the presence of a similar parameter estimation mode. To aide in this effort we restrict the prior on the chirp mass. We also only consider the results for the detector frame masses, as the luminosity distance estimate is poor for single detector analysis due to the lack of information on the sky position. We will only consider the triggers that might contain a sub-solar mass component.

\begin{figure}[tb]
        \centering
    	\includegraphics[width=\columnwidth]{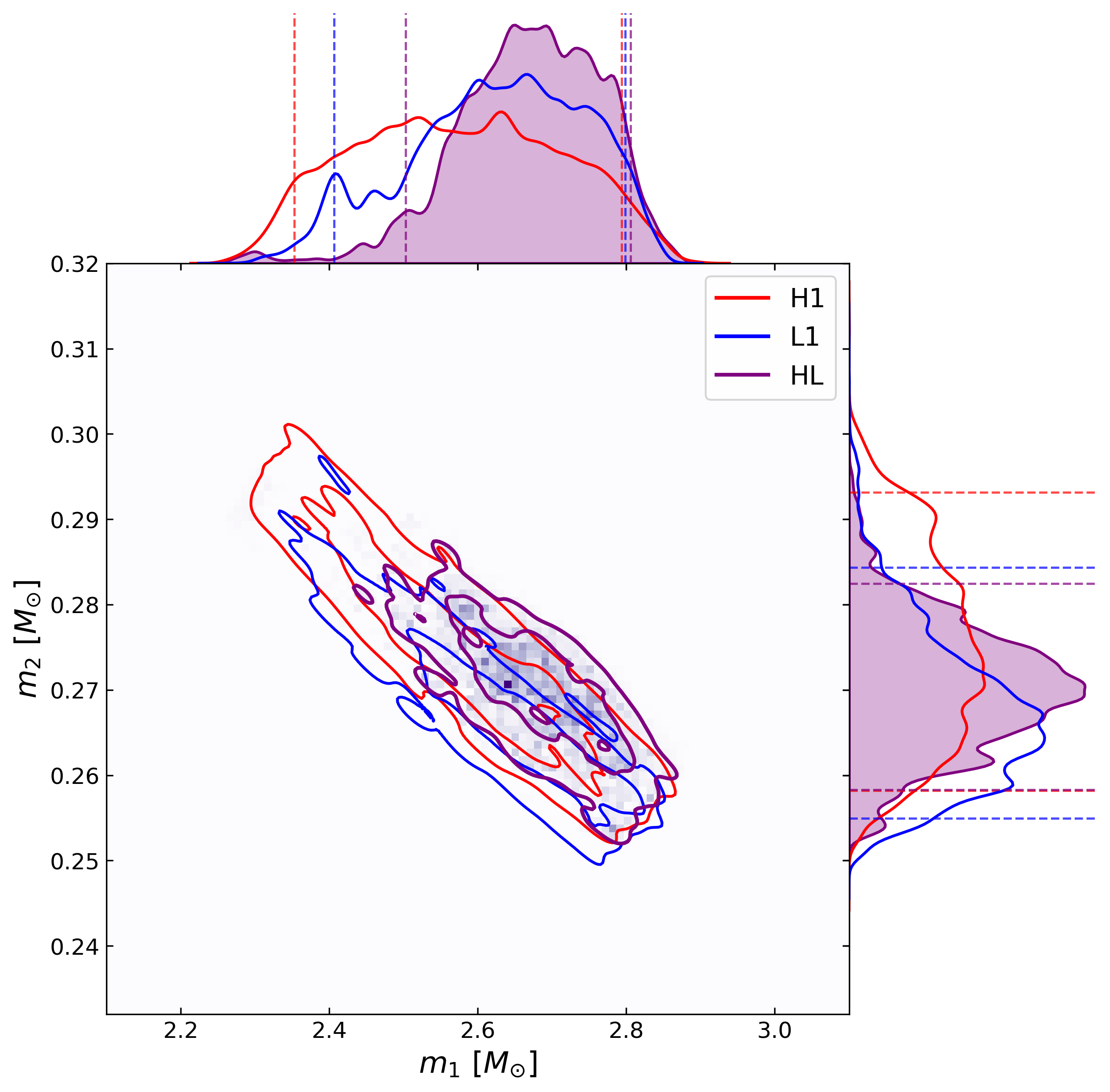}
    	\caption{Posterior distributions for the detector frame component masses, $m_1$ and $m_2$, for the trigger at 2023-09-14 21:02:19.50 UTC. The estimates come from analyzing only the data from L1, only from H1, and analyzing H1 and L1 together.}
        \label{fig:1378760557_256_1s}
    \end{figure}

\begin{figure}[tb]
        \centering
    	\includegraphics[width=\columnwidth]{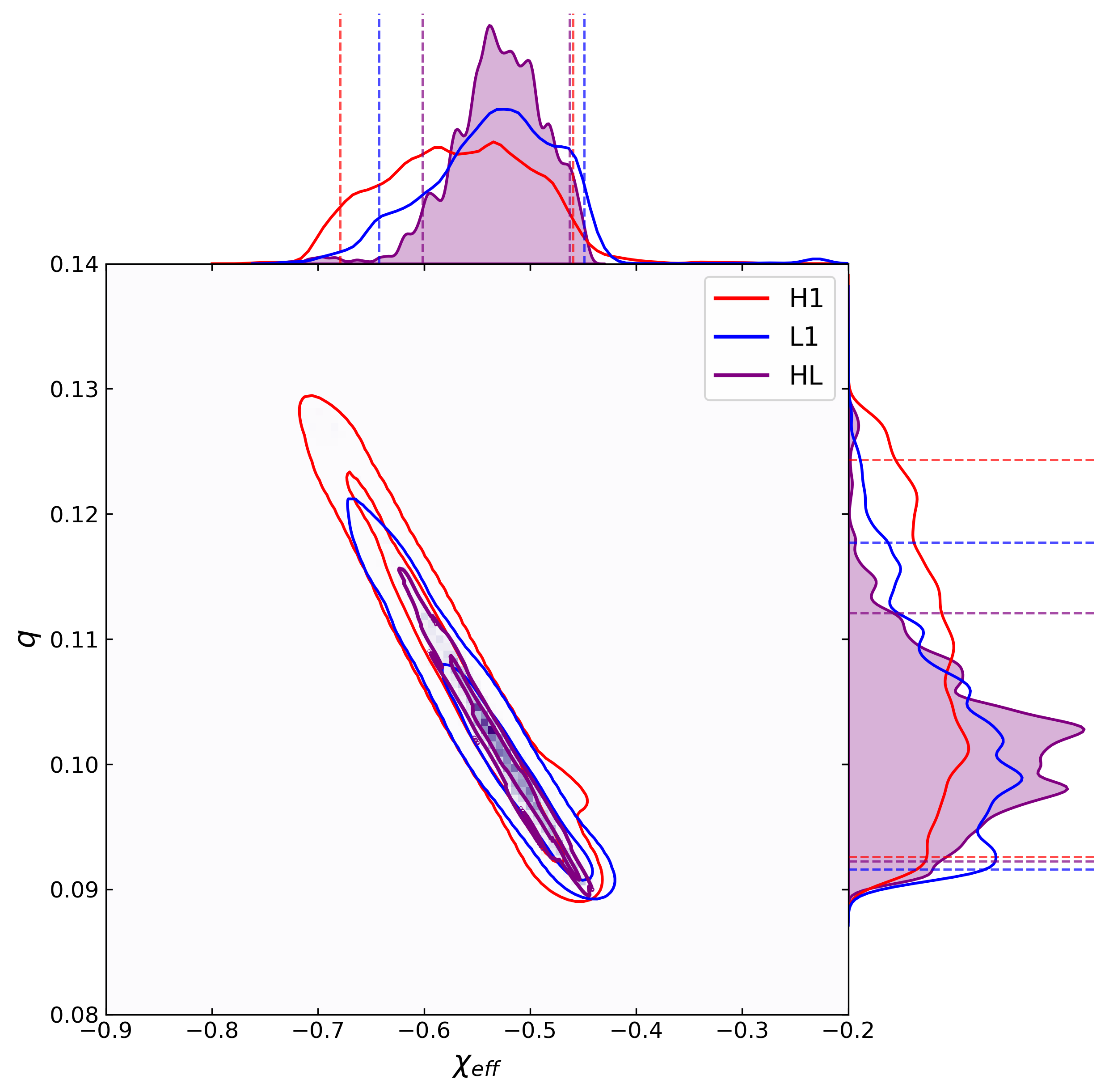}
    	\caption{Posterior distributions for the effective spin $\chi_{\rm eff}$ and the mass ratio $q=m_2/m_1$ for the trigger at 2023-09-14 21:02:19.50 UTC. The estimates come from analyzing only the data from L1, only from H1, and analyzing H1 and L1 together.}
        \label{fig:chi-q-1378760557_256_1s}
    \end{figure}

We first consider the trigger at 2023-09-14 21:02:19.50 UTC. For the analysis of the data from H1 alone, and L1 alone, we set the prior range for the chirp mass to be $0.67 ~ \textup{M}_\odot < m_c < 0.69 ~ \textup{M}_\odot$. The parameter estimation results are presented in Table~\ref{tab:LIGO_1D} and Fig.~\ref{fig:1378760557_256_1s}. For this trigger we also compare the parameter estimation results for the effective spin $\chi_{\rm eff}$ and the mass ratio $q=m_2/m_1$; see Fig.~\ref{fig:chi-q-1378760557_256_1s}. As can be seen, the parameter estimates using single detector (H1 and L1) data are consistent with the results on the combined H1-L1 data.

\begin{table*}
\caption{\label{tab:LIGO_1D} Presented are the parameter estimation results for the LIGO trigger at 2023-09-14 21:02:19.50 UTC. The data from only one detector are analyzed. The chirp mass $m_c$ and the component masses $m_1$ and $m_2$ are in the detector frame.}
\begin{ruledtabular}
\begin{tabular}{cccccccc}
 Detector& $m_{c}$ ($\textup{M}_\odot$)& $m_{1}$ ($\textup{M}_\odot$)&$m_{2}$ ($\textup{M}_\odot$)&$\chi_{\textup{eff}}$&$D_L$ (Mpc)&SNR&Data length (s) \\  \hline
H1&$0.682^{+0.001}_{-0.001}$&$2.75^{+0.18}_{-0.16}$&$0.26^{+0.01}_{-0.01}$&$-0.49^{+0.09}_{-0.06}$&$182.16^{+62.10}_{-49.31}$ &6.46 & 256\\
L1&$0.684^{+0.001}_{-0.001}$&$3.01^{+0.42}_{-0.55}$&$0.24^{+0.03}_{-0.02}$&$-0.28^{+0.12}_{-0.23}$&$216.80^{+276.51}_{-59.80}$ & 6.44&256
\end{tabular}
\end{ruledtabular}
\end{table*}

For the trigger at 2023-07-19 11:50:50.27 UTC the parameter estimation results are presented in Table~\ref{tab:LIGO_1D2}, and Fig.~\ref{fig:1373802668_256_1s} in the Appendix. For the analysis of the data from H1 alone the prior range for the chirp mass is $0.33 ~ \textup{M}_\odot < m_c < 0.38 ~ \textup{M}_\odot$. The parameter estimates are in good agreement with those from the combined H1-L1 data. For the analysis of the data from L1 alone the prior range for the chirp mass is $0.34 ~ \textup{M}_\odot < m_c < 0.38 ~ \textup{M}_\odot$. The estimate of the masses are in decent agreement with that from the combined H1-L1 data.

\begin{table*}
\caption{\label{tab:LIGO_1D2} Presented are the parameter estimation results for the LIGO trigger at 2023-07-19 11:50:50.27 UTC. The data from only one detector are analyzed. The chirp mass $m_c$ is for the detector frame, while the component masses $m_1$ and $m_2$ are from the source frame.}
\begin{ruledtabular}
\begin{tabular}{cccccccc}
 Detector& $m_{c}$ ($\textup{M}_\odot$)& $m_{1}$ ($\textup{M}_\odot$)&$m_{2}$ ($\textup{M}_\odot$)&$\chi_{\textup{eff}}$&$D_L$ (Mpc)&SNR&Data length (s) \\  \hline
H1&$0.3609^{+0.0001}_{-0.0001}$&$0.7903^{+0.0776}_{-0.0388}$&$0.2246^{+0.0089}_{-0.0157}$&$-0.1102^{+0.0511}_{-0.0320}$&$98.18^{+23.88}_{-28.93}$ &7.13 & 512\\
L1&$0.3674^{+0.0001}_{-0.0028}$&$0.7837^{+0.1494}_{-0.1341}$&$0.2299^{+0.0324}_{-0.0415}$&$0.5522^{+0.0232}_{-0.4389}$&$125.35^{+547.03}_{-29.12}$ &6.08& 256
\end{tabular}
\end{ruledtabular}
\end{table*}


For the trigger at 2023-10-14 08:15:06.33 UTC, the single-detector parameter estimation results are presented in Table~\ref{tab:LIGO_1D3} and Fig.~\ref{fig:1381306524_256_1s}. For these analyses, the chirp mass prior was restricted to $0.50\,M_{\odot} < m_{c} < 0.60\,M_{\odot}$. While the L1 estimate ($m_c = 0.55678_{-0.00027}^{+0.00058}\,M_\odot$) agrees closely with the combined H1-L1 result ($m_c = 0.55659_{-0.00036}^{+0.00040}\,M_\odot$), the H1 estimate ($m_c = 0.54412_{-0.00036}^{+0.00056}\,M_\odot$) exhibits a noticeable shift. Crucially, the 90\% credible intervals for $m_c$ between H1 and L1 do not overlap. Such single-detector parameter tension at low SNR ($\sim 6.4$) is a primary diagnostic that the trigger is non-astrophysical, resulting from uncorrelated noise fluctuations or instrumental transients in the two detectors rather than a coherent gravitational-wave signal.

\begin{table*}
\caption{\label{tab:LIGO_1D3} Presented are the parameter estimation results for the LIGO trigger at 2023-10-14 08:15:06.33 UTC. The data from only one detector are analyzed. The chirp mass $m_c$ is for the detector frame, while the component masses $m_1$ and $m_2$ are from the source frame.}
\begin{ruledtabular}
\begin{tabular}{cccccccc}
 Detector& $m_{c}$ ($\textup{M}_\odot$)& $m_{1}$ ($\textup{M}_\odot$)&$m_{2}$ ($\textup{M}_\odot$)&$\chi_{\textup{eff}}$&$D_L$ (Mpc)&SNR&Data length (s) \\  \hline
H1&$0.54412^{+0.00056}_{-0.00036}$&$1.28692^{+0.43646}_{-0.37464}$&$0.30869^{+0.09476}_{-0.06099}$&$0.65475^{+0.02541}_{-0.03674}$&$168.15^{+73.18}_{-46.57}$ &6.43 & 128\\
L1&$0.55678^{+0.00058}_{-0.00027}$&$2.25840^{+0.38876}_{-0.23308}$&$0.21122^{+0.01620}_{-0.02359}$&$0.78831^{+0.01382}_{-0.01696}$&$179.18^{+75.45}_{-44.21}$ &6.42& 128
\end{tabular}
\end{ruledtabular}
\end{table*}

\section{Data Quality Checks}\label{sec:dq}

The parameter estimation analyses operate under the assumption that the instrument noise is stationary and Gaussian, so it is important to check if those assumptions hold. More precisely, the search algorithms are robust against certain types of non-stationarity and non-Gaussianity, so that particularly bad stretches of data are unlikely to yield significant triggers. In addition, the QuickCBC algorithm we used to perform Bayesian inference automatically subtracts noise transients (glitches) using wavelet denoising, so the results should be fairly robust against instrument glitches. However, the likelihood function assumes the noise is wide sense stationary once glitches have been removed, and this may not be true when analyzing long stretches of data.

We applied multiple data quality tests based on time-frequency transforms of the whitened strain data. A new version of the BayesWave algorithm~\cite{Cornish_2015,PhysRevD.91.084034,Cornish:2020dwh, PhysRevD.109.064040} was used to whiten the data and subtract spectral lines. The spectral lines are coherent, almost sinusoidal features introduced by the suspension system and couplings to the electric grid. Continuous, non-orthogonal wavelet transforms, such as ``Q-scans'', provide qualitative checks on the data quality, while orthogonal wavelet transforms, such as the Wilson-Daubechies-Meyer (WDM) wavelet wavepackets, can be used to perform quantitative data quality tests. We experimented with several different tests, and in the end we found three WDM based tests to be the most informative. In each case the starting point was the whitened, line subtracted strain data. This data was then WDM transformed using pixels with bandwidth $\Delta F = 4$ Hz and duration $\Delta T = 0.125$ seconds. The transform was normalized to have zero mean and unit variance for $N(0,1)$ input data.

\begin{figure}[tb]
        \centering
    	\includegraphics[width=\columnwidth]{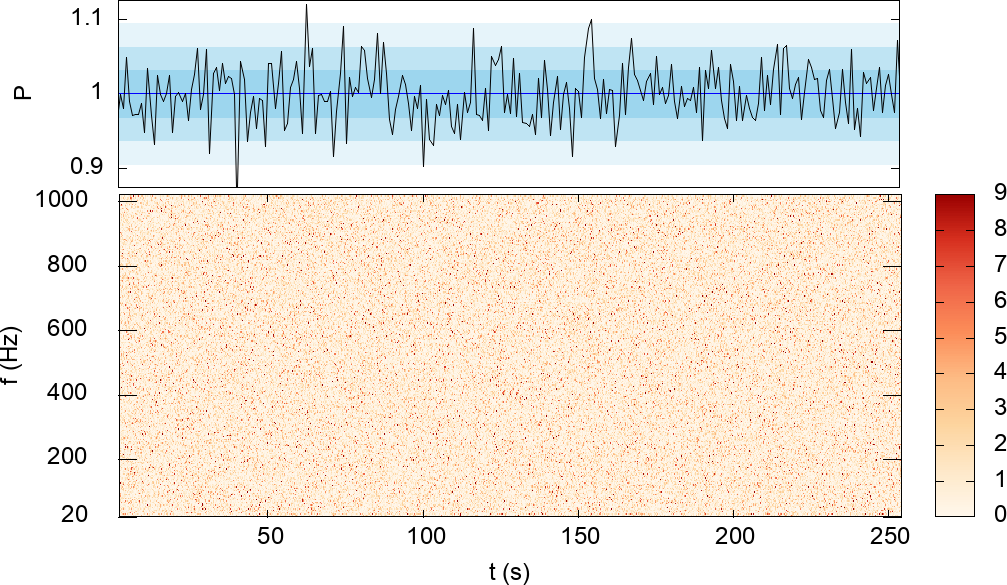}
    	\caption{The WDM power distribution for whitened data from the L1 detector for the trigger from  2023-09-14 21:02:19.50 UTC (1378760557.50 GPS). The data covers the 256 seconds from GPS time 1378760305 to 1378760561. The power level stays within the expected range, indicating that the data are stationary during this period. Also, there are no visual indication of noise transients.}
        \label{fig:power_1378760557_1}
    \end{figure}

The first diagnostic is to plot the squared amplitude of the WDM transform. We examine the data around the trigger at trigger from  2023-09-14 21:02:19.50 UTC; see Fig.~\ref{fig:power_1378760557_1}. The pixels in the transform should follow a $\chi^2$ distribution with one degree of freedom. To test for stationarity, the power is summed across frequencies for each time. The summed power should have mean 1 and variance $1/N_f$, where $N_f$ are the number of frequency pixels that are summed over.

\begin{figure}[tb]
        \centering
    	\includegraphics[width=\columnwidth]{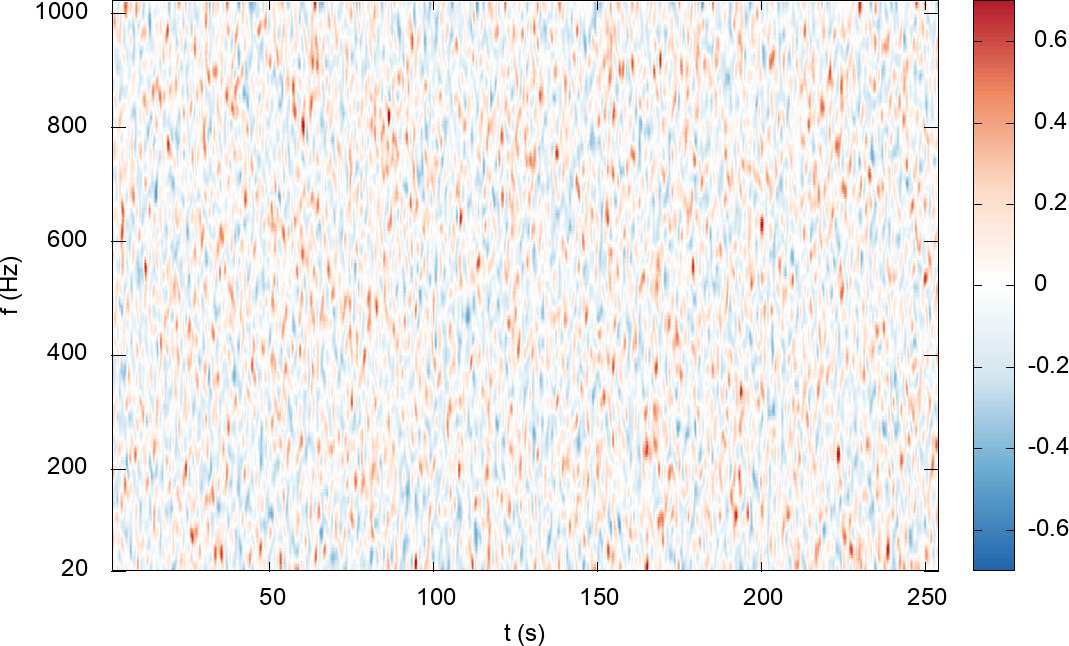}
    	\caption{The smoothed WDM power fluctuation map for the same stretch of data shown in Fig.~\ref{fig:power_1378760557_1}.}
        \label{fig:tranPS_1369307270_1}
    \end{figure}

The second diagnostic is similar, and but more qualitative; see Fig.~\ref{fig:tranPS_1369307270_1}. A Gaussian smoothing window with $\sigma_f = 3 \Delta F$ and $\sigma_t = 3 \Delta T$ is applied to the WDM power map and the expected mean value of unity is subtracted. The resulting smoothed power fluctuation map is useful for catching more localized departures from stationarity.

\begin{figure}[tb]
        \centering
    	\includegraphics[width=\columnwidth]{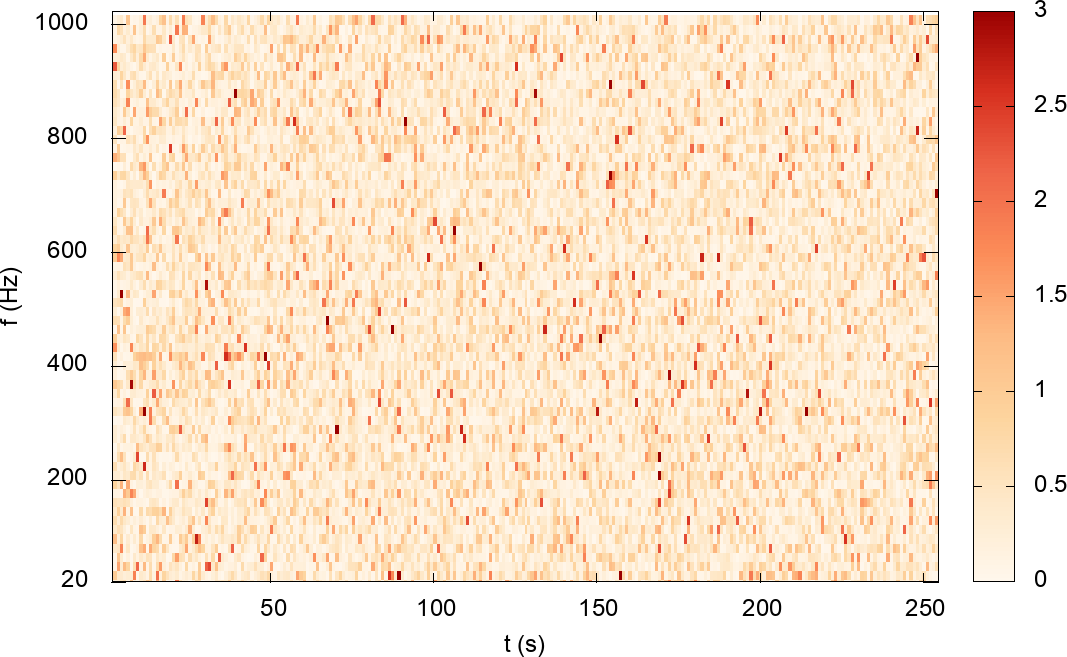}
    	\caption{Anderson-Darling p-values for the same stretch of data shown in Fig.~\ref{fig:power_1378760557_1}. The color scale shows $-\log_{10}(p)$. }
        \label{fig:pvalues_1369307270_1}
    \end{figure}

Finally, we apply a quantitative test of stationarity and Gaussianity. The test takes the original WDM transform and applies the Anderson-Darling test to data in $n_f \times n_t$ blocks of pixels. We apply the variant of the Anderson-Darling test that assumes the mean and variance are known. That way the test catches both local excesses in the power and deviations from a normal distribution. The Anderson-Darling statistic for this test is then converted to a p-value that measures consistency with the hypothesis that the WDM pixels are $N(0,1)$ distributed. See Fig.~\ref{fig:pvalues_1369307270_1}.

\begin{figure}[tb]
        \centering
    	\includegraphics[width=\columnwidth]{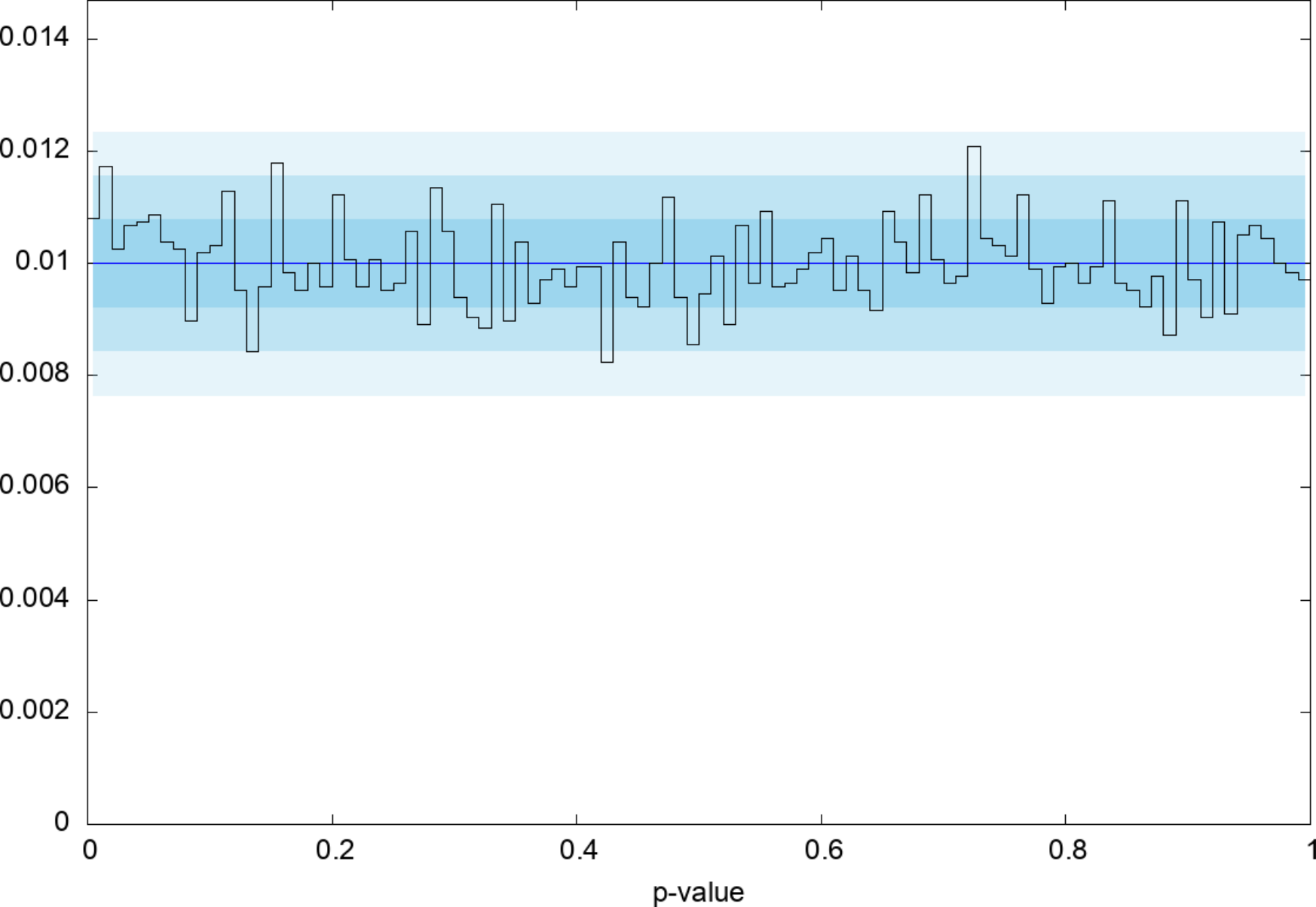}
    	\caption{Anderson-Darling p-value histogram for the same stretch of data shown in Fig.~\ref{fig:power_1378760557_1}.}
        \label{fig:phist_1369307270_1}
    \end{figure}

A histogram of the p-values, shown with the expected 1-2-3 sigma sample variation, confirms that this stretch of data is consistent with our assumption of stationary, Gaussian noise. See Fig.~\ref{fig:phist_1369307270_1}.

We found that all three tests were similarly effective at flagging bad data. The smoothed power fluctuation maps provided the best quick visual check, while the p-values were useful in measuring just how significant the deviations were. Overall, the majority of the sub-solar mass candidates were found in clean, stationary, glitch free data. The absence of significant glitches in most of the 256 second long data stretches is somewhat surprising given that there are typically 2-3 glitches with ${\rm SNR} > 10$ every 1000 seconds. A few of the sub-solar mass candidates were found in data that had mild to moderate departures from stationarity or Gaussianity in one or more detector. 

\begin{figure}[tb]
        \centering
    	\includegraphics[width=\columnwidth]{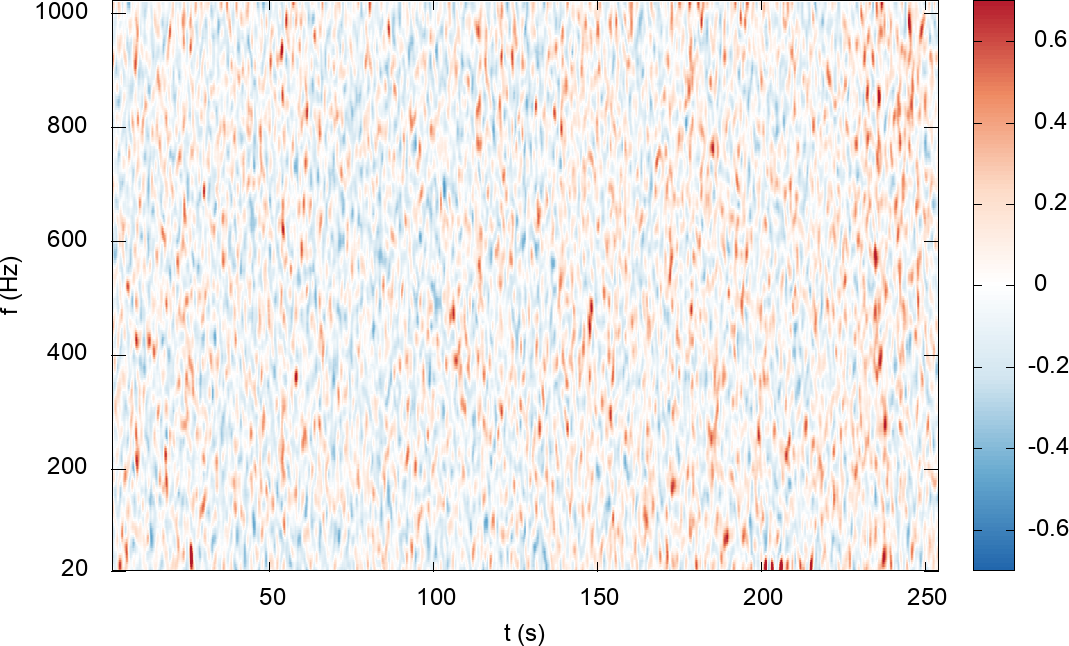}
    	\includegraphics[width=\columnwidth]{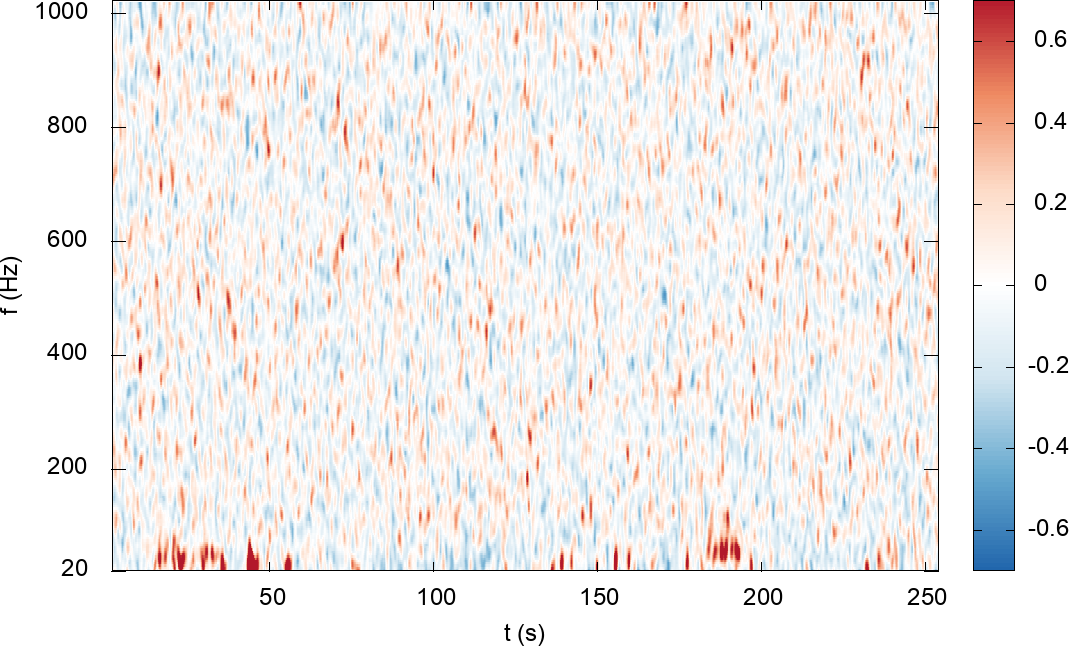}
    	\caption{Data with sub-solar mass candidates that had some data quality issues. L1 data for trigger  2020-03-08 18:05:53.02 UTC (1267725971.02 GPS) - slightly non-stationary near the merger (above). Trigger 2023-07-19 11:50:50.27 UTC (1373802668.27 GPS) - low frequency glitching in H1 (below).}
        \label{fig:bad-data1}
    \end{figure}
    
\begin{figure}[tb]
        \centering
    	\includegraphics[width=\columnwidth]{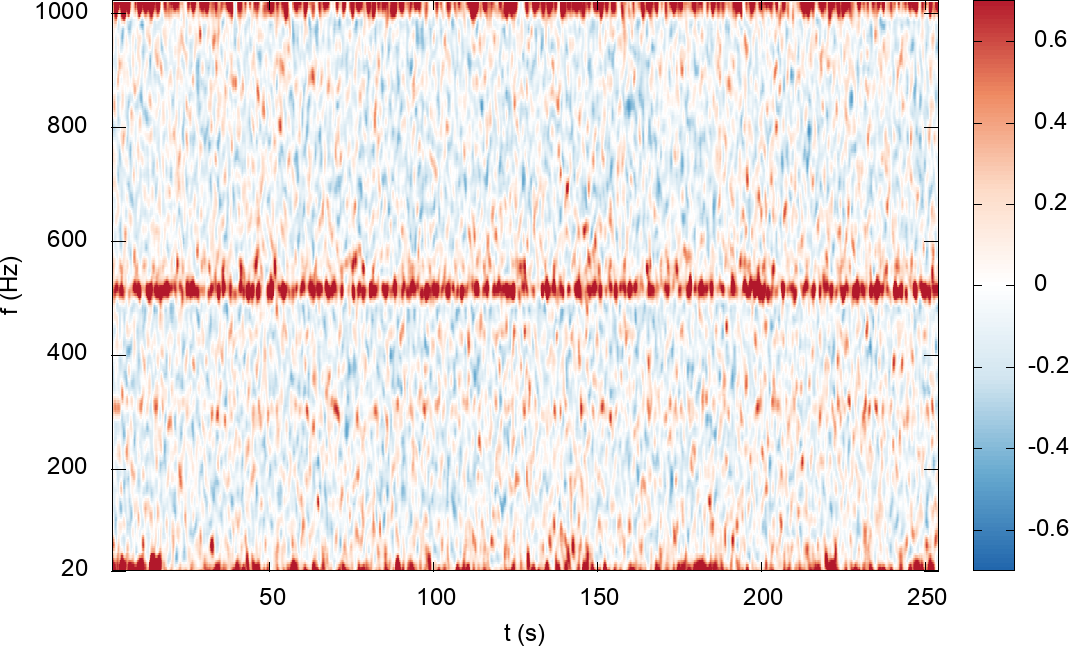}
    	\includegraphics[width=\columnwidth]{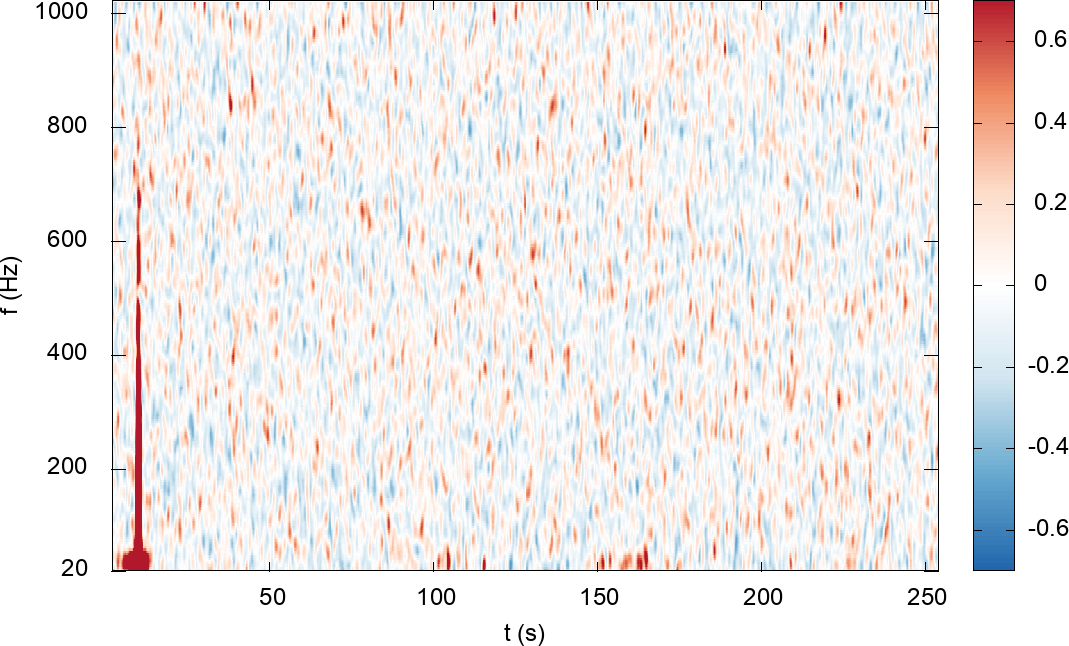}
    	\caption{Data with sub-solar mass candidates that had some data quality issues.  For the trigger at 2023-07-19 11:50:50.27 UTC (1373802668.27 GPS) there are non-stationary lines in L1 (above).  For trigger 2023-08-10 10:10:03.37 UTC (1375697421.37 GPS), there is a loud glitch in H1 (below).}
        \label{fig:bad-data2}
    \end{figure}    

We do not expect that the glitches would have much impact given that the analysis is performed on glitch-cleaned data. More concerning is the mild non-stationarity in L1 for the trigger at 2020-03-08 18:05:53.02 UTC (1267725971.02 GPS), and the significant non-stationarity around spectral lines in H1 for the trigger at 2023-07-19 11:50:50.27 UTC (1373802668.27 GPS). See Fig.~\ref{fig:bad-data1}. Further data quality issues can be seen at L1 for 2023-07-19 11:50:50.27 UTC (1373802668.27 GPS) with noise lines, and a loud glitch with the H1 data for the trigger at 2023-08-10 10:10:03.37 UTC (1375697421.37 GPS). See Fig.~\ref{fig:bad-data2}.

\section{Signal injection study}\label{sec:inj}
\begin{figure*}[tb]
        \centering
        \includegraphics[width=\linewidth]{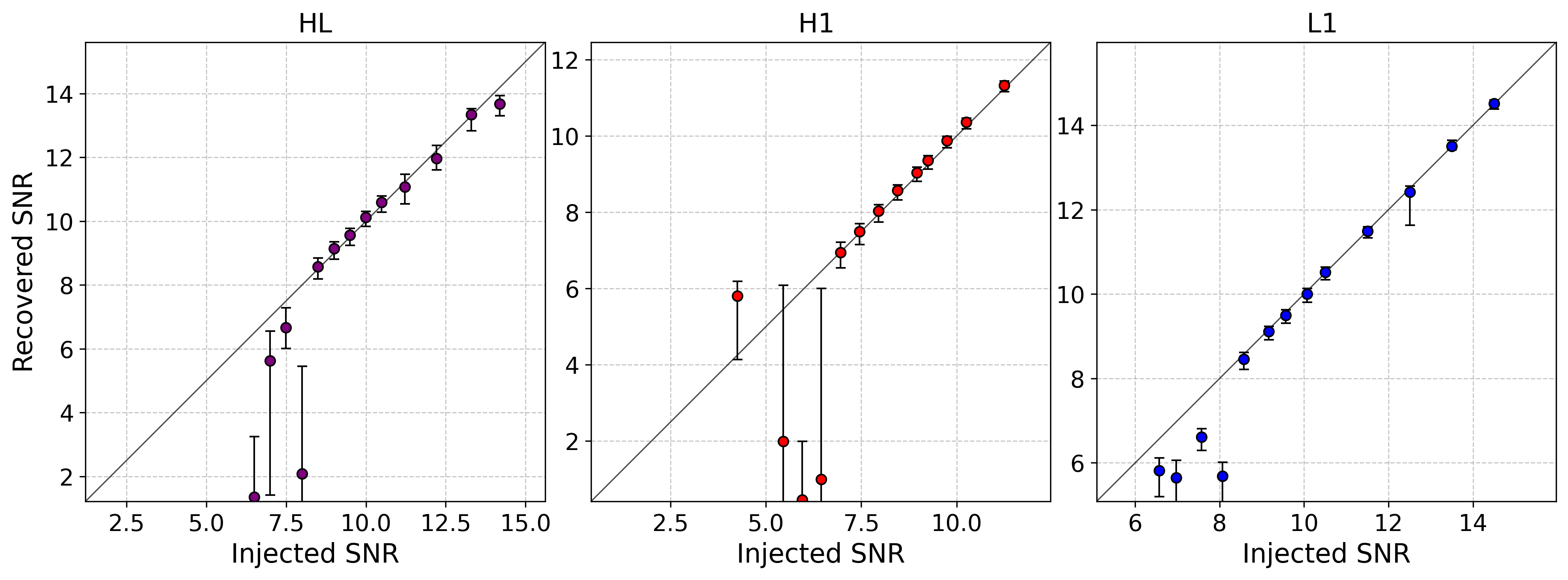}
    	\caption{Recovered SNR as a function of the injected SNR for the combined network of detectors, and for each detector individually. The solid diagonal line indicates the ideal case of perfect recovery. Injections use the parameters of the 2023‑09‑14 21:02:19.50 UTC trigger. Note that an SNR $\sim 8$ or larger is needed for the parameter estimation to accurately reproduce the injected parameters.}
        \label{fig:injectionSNR}
    \end{figure*}

In order to better understand when compact binary coalescence signals may be identified in LIGO O4a data, we conducted a signal injection study. We mimic a signal that may contain a sub solar mass component, specifically the  2023-09-14 21:02:19.50 UTC trigger. We use the H1 and L1 data, but inject a signal where the merger happens 2000s after the original trigger merger time. The time shift was chosen to mimic the local background noise characteristics (PSD), of the real trigger without risking signal overlap or contamination. The injection parameters match the trigger~\cite{LIGOScientific:2026XXX}; $m_1 = 2.63 ~ \textup{M}_\odot$, $m_2 = 0.29 ~ \textup{M}_\odot$, $\chi_1 = -0.72$, and $\chi_2 = -0.02$. The injected signal-to-noise value is varied in order to understand which value allows to retrieve parameters consistent with the injected values. Figure ~\ref{fig:injectionSNR} shows the recovered SNR as a function of the injected SNR for the combined network (HL) as well as the individual detectors (H1 and L1). When individual detector data, H1 and L1, are used, a signal-to-noise ratio of $\sim 7$ and $\sim 8.5$ correspondingly are required to get acceptable parameter estimation results, consistent with the injection. When the combined H1-L1 data are used, a network signal-to-noise ratio of $\sim 8.2$ is needed to accurately recover the injection parameters, which corresponds to signal-to-noise ratios of $\sim 5.7$ and $\sim 6.2$ in each individual detector correspondingly. In both cases, for all the injections above these SNR thresholds, the true injected values are contained within the resulting posterior distribution.

\section{Conclusions}\label{sec:conclusions}
We have conducted parameter estimation on the seven potential sub-threshold compact binary coalescence triggers from LIGO data in observing run O4a, as reported by the LVK~\cite{LIGOScientific:2026XXX}. Of the seven, with parameter estimation we can reproduce mass and spin parameters that are consistent with those reported from the signal searches. Three of these triggers could correspond to compact binary systems containing at least one sub-solar mass component. The other two triggers, if real, would likely correspond to neutron star - black hole binaries.

We have also conducted parameter estimation on three triggers reported by groups outside of the LVK~\cite{Niu:2025nha,Kacanja:2026byy}. We have also analyzed a LIGO-Virgo trigger from O3~\cite{Prunier:2023uoo}. These triggers were identified by different signal search pipelines, but none of these events were significant enough to be declared a detection. Of the four triggers studied, we have identified three that may contain a sub-solar mass object. The parameter estimation results are presented in Appendix~\ref{App:others}.

If confirmed, compact binaries containing sub-solar mass black holes would have profound implications. Such objects are not expected from ordinary stellar evolution, and their observation would therefore constitute one of the cleanest gravitational-wave signatures of a primordial black-hole population. The candidate systems identified here are not statistically significant detections, so no claim of a PBH discovery can be made. Nevertheless, our results show that several of the most significant sub-solar-mass triggers in O4a are compatible, at the level of Bayesian parameter estimation, with compact binaries containing objects in precisely the mass range where a primordial interpretation is most compelling.

If any of these sub-threshold triggers is real, then we should see similar events at high significance as the detector sensitivities improve and more data is taken.

As noted in our study, parameter estimation on a potential sub-solar mass compact binary coalescence triggers is challenging. With low chirp masses, the required signal length for analysis can be very long in duration. In our study we have used the QuickCBC parameter estimation code~\cite{Cornish:2021wxy}, but computing memory requirements can still be an issue. The long data stretches also raise data quality issues, especially with the appearance of glitches. Non-Gaussian data can also be a problem. We have displayed different data quality tools that can help in assuring that the possible events happened in good data stretches. Data quality issues were identified of the triggers at  2020-03-08 18:05:53.02 UTC, 2023-07-19 11:50:50.27 UTC, and 2023-08-10 10:10:03.37 UTC. When the mass ratio or the chirp mass are small, there can be issues with the signal waveforms. In our study here we have studied triggers with detector frame chirp masses as small as $m_c \sim 0.3 ~ \textup{M}_\odot$ and mass ratios as low as $q = m_2/m_1 \sim 0.1$. 

Additionally, our analysis utilized the IMRPhenomD waveform model, which incorporates only the dominant $(\ell, m) = (2,2)$ spherical harmonic mode. However, several candidate triggers investigated here feature pronounced mass asymmetries, most notably the trigger at 2023-09-14 21:02:19.50 UTC, where $m_1 \approx 2.62\,M_{\odot}$ and $m_2 \approx 0.27\,M_{\odot}$ ($q \approx 0.1$). For systems with such extreme mass ratios, higher-order multipole modes (such as $(\ell, m) = (3,3)$ and $(4,4)$) can become significantly excited, offering crucial leverage to break degeneracies between mass ratio, inclination, and luminosity distance. Implementing waveform models with higher-order modes (IMRPhenomXHM or SEOBNRv5HM) represents a key future enhancement for sub-solar mass parameter estimation, especially as higher SNR candidates emerge in upcoming observing runs.

A signal injection study was also done. In order to have the parameter estimation correctly return the injected signal parameters, a network SNR of $\sim 8$ is needed. When running on combined LIGO H1 and L1 data, this implies an individual SNR of $\sim 6$. However, for a single detector analysis, one needs an SNR of $\sim 8$. In analyzing the triggers we have looked for consistency in the parameter estimation results when using the combined H1-L1 data, and then using the H1 data alone, and the L1 data alone. When analysing data from a single detector we have had to narrow the prior on the chirp mass.

We have also examined how the SNR in each detector grows with frequency, and how this compares to the theoretical expectation. For the triggers we analyzed where there may be a sub-solar mass component, we have seen that the SNR grows as expected.

From the PBH perspective, the next step is not only to improve the detection statistic for individual events, but also to build a population-level picture. A confirmed population of sub-solar compact binaries would allow one to constrain the PBH abundance, mass function, spin distribution and binary-formation channel. Conversely, the absence of convincing detections in future observing runs will place increasingly strong limits on PBHs in the sub-solar range. Future analyses should therefore combine parameter estimation, data-quality studies, searches for tidal effects, and population modelling in order to distinguish PBHs from neutron stars or other exotic compact objects.

Our analysis does not give an improved detection statistic. However, we have used our parameter estimation and data quality methods to show that many of the triggers are in clean data, and produce consistent results in each of the LIGO detectors. We can not rule out the possibility that some of these events are real signals produced by compact binary systems where at least one mass component is sub-solar mass. 

\begin{acknowledgments}
This material is based upon work supported by NSF’s LIGO Laboratory which is a major facility fully funded by the National Science Foundation. This research has made use of data~\cite{KAGRA:2023pio,LIGOScientific:2025snk} or software obtained from the Gravitational Wave Open Science Center (gwosc.org), a service of the LIGO Scientific Collaboration, the Virgo Collaboration, and KAGRA. This material is based upon work supported by NSF's LIGO Laboratory which is a major facility fully funded by the National Science Foundation, as well as the Science and Technology Facilities Council (STFC) of the United Kingdom, the Max-Planck-Society (MPS), and the State of Niedersachsen/Germany for support of the construction of Advanced LIGO and construction and operation of the GEO600 detector. Additional support for Advanced LIGO was provided by the Australian Research Council. Virgo is funded, through the European Gravitational Observatory (EGO), by the French Centre National de Recherche Scientifique (CNRS), the Italian Istituto Nazionale di Fisica Nucleare (INFN) and the Dutch Nikhef, with contributions by institutions from Belgium, Germany, Greece, Hungary, Ireland, Japan, Monaco, Poland, Portugal, Spain. KAGRA is supported by Ministry of Education, Culture, Sports, Science and Technology (MEXT), Japan Society for the Promotion of Science (JSPS) in Japan; National Research Foundation (NRF) and Ministry of Science and ICT (MSIT) in Korea; Academia Sinica (AS) and National Science and Technology Council (NSTC) in Taiwan.

The authors are grateful for computational resources provided by the LIGO Laboratory and supported by the National Science Foundation Grants PHY-0757058 and PHY-0823459. Neil Cornish acknowledges NSF award PHY2513363 and Simons Foundation award SFI-MPS-BH-00012593-04. Mairi Sakellariadou acknowledges support from the Science and Technology Facility Council (STFC), UK, under the research grant ST/X000753/1. This work was supported by the French government through the France 2030 investment plan managed by the National Research Agency (ANR), as part of the Initiative of Excellence of Université Côte d’Azur under reference number ANR-15-IDEX-01. We thank Gregory Ashton, Sasha Kashlinsky and Viola Sordini for their comments. This manuscript was assigned LIGO-Document number LIGO-P2600340.
\end{acknowledgments}

\begin{appendix}

\section{Other events}\label{sub:S231109ci}
\label{App:others}

In this Appendix, we extend our analysis to additional candidate triggers identified by non-LVK searches and from a previous LVK sub-solar mass search. The inclusion of these candidates works to further validate our methodology, ensuring that the performance of our framework is consistent across different search pipelines, candidate origins, and observing runs. These candidates are presented in Table \ref{tab:table_othertriggers}.

\begin{table*}
\caption{\label{tab:table_othertriggers} The table includes two triggers from independent sub-solar mass searches (2023-11-09 23:54:56.048 UTC~\cite{Niu:2025nha} and 2023-08-14 00:43:53.584 UTC~\cite{Kacanja:2026byy}), alongside one trigger from the LVK O3b sub-solar mass search that was then studied via parameter estimation (2020-03-08 18:05:53.02~\cite{LVK:2022ydq}).}
\begin{ruledtabular}
\begin{tabular}{ccccccc}
 Time&Primary mass $m_{1}$&Secondary mass $m_{2}$&Chirp mass&Detectors
&Network SNR&FAR \\ 
(UTC)& Detector Frame ($\textup{M}_\odot$)&Detector Frame ($\textup{M}_\odot$)&Detector Frame ($\textup{M}_\odot$) & & & yr$^{-1}$ \\ \hline
 2023-11-09 23:54:56.048&1.78&1.27&1.30&HL&9.62&0.02\\
 2023-08-14 00:43:53.584&[0.1, 2]&[0.1,1]&0.30&HL&9.66&10.02\\
 2020-03-08 18:05:53.02&0.78&0.23&0.35&HL&8.90&0.20
\end{tabular}
\end{ruledtabular}
\end{table*}

We first consider the LVK sub-threshold trigger GW231109\_235456 involving H1 and L1 at 2023-11-09 23:54:56.048 UTC. This trigger was reported to be a possible binary neutron star merger~\cite{Niu:2025nha}. This study reported a network signal-to-noise ratio of 9.7, and using a low spin prior, a detector frame chirp mass of $m_c = 1.3063^{+0.0003}_{-0.0003} ~ \textup{M}_\odot$ and an effective spin of $\chi_{\rm eff} = 0.014^{+0.016}_{-0.018}$. With a luminosity distance estimate of $D_L = 169^{+72}_{-70}$ Mpc, the source frame component masses were estimated to fall in the ranges $1.40 < m_1 < 1.65 ~ \textup{M}_\odot$ and $1.27 < m_2 < 1.49 ~ \textup{M}_\odot$, hence a possible binary neutron star system. 

Another group has searched the LIGO O4a data in search of compact binary mergers involving at least one sub solar mass component~\cite{Kacanja:2026byy}. No convincing detection was made, but two top candidates were reported. One was at 2023-05-31 16:22:47.539 UTC, with a detector frame chirp mass of $m_c = 0.32 ~ \textup{M}_\odot$ and a network signal-to-noise ratio of 8.98. The other candidate was at 2023-08-14 00:43:53.584 UTC, with a detector frame chirp mass of $m_c = 0.30 ~ \textup{M}_\odot$ and a network signal-to-noise ratio of 9.66. Both triggers involved data from both H1 and L1.
The parameter estimation results for the trigger at 2023-05-31 16:22:47.539 UTC were inconsistent and not believable; we do not address the trigger further.

The potential SSM200308 trigger occurred at 2020-03-08 18:05:53.02 UTC was detected by the LVK with the GSTLAL pipeline with a false alarm rate of 0.2 yr$^{-1}$~\cite{LVK:2022ydq}, a network signal-to-noise ratio of 8.90, and a detector frame chirp mass of $m_c = 0.355 ~ \textup{M}_\odot$. This trigger was subsequently studied with parameter estimation, with results presented in~\cite{Prunier:2023uoo}. 

The parameter estimation results for the triggers at 2023-11-09 23:54:56.048 UTC, 2023-08-14 00:43:53.584 UTC, and 2020-03-08 18:05:53.02 UTC are presented in Tables~\ref{tab:table_PE_others} and \ref{tab:mc_and_chi_others}. 

For the trigger at 2023-11-09 23:54:56.048 UTC, if real, would be produced by a binary neutron star. However, the smaller compact object does have some probability to be close to $1.0 ~ \textup{M}_\odot$. See Fig.~\ref{fig:1383609314_128_2} in the Appendix. As shown in Fig.~\ref{fig:1383609314_SNR2_H1L1_together}  the accumulation of SNR$^2(f)$ is consistent with the expected CBC signal with masses given in Table~\ref{tab:table_PE_others} for the trigger at 2023-11-09 23:54:56.048 UTC.

For the trigger at 2023-08-14 00:43:53.584 UTC, if real, would be produced by a binary system containing two sub-solar mass objects. There is a small probability that the primary mass could exceed 1.0.  See Fig.~\ref{fig:1376009051_512_2} in the Appendix. As shown in Fig.~\ref{fig:1376009051_SNR2_H1L1_together},  the accumulation of SNR$^2(f)$ is consistent with the expected CBC signal with masses given in Table~\ref{tab:table_PE_others} for the trigger at 2023-08-14 00:43:53.584 UTC.

For the trigger at 2020-03-08 18:05:53.02 UTC, if real, would be produced by a binary system containing two sub-solar mass objects. See Fig.~\ref{fig:1267725971_256_2} in the Appendix. As shown in Fig.~\ref{fig:1267725971_SNR2_H1L1V1_together}, the accumulation of SNR$^2(f)$ is consistent with the expected CBC signal with masses given in Table~\ref{tab:table_PE_others} for the trigger at 2020-03-08 18:05:53.02 UTC. Our recovered parameters are in good agreement with the independent parameter estimation study previously performed by Prunier et al.~\cite{Prunier:2023uoo}, who reported component masses of $m_{1} = 0.62_{-0.20}^{+0.46} \mathcal{M}_{\odot}$ and $m_{2} = 0.27_{-0.12}^{+0.10} \mathcal{M}_{\odot}$. This validates the consistency of the parameters recovery across different frameworks. We note that for this trigger, when we attempted parameter estimation on H1 data alone, or L1 data alone, we were not able to have acceptable results with any reasonable SNR value.

\begin{table*}
\caption{\label{tab:table_PE_others} Parameter estimation results for the triggers at 2023-11-09 23:54:56.048 UTC~\cite{Niu:2025nha}, 2023-08-14 00:43:53.584 UTC~\cite{Kacanja:2026byy}, and 2020-03-08 18:05:53.02~\cite{LVK:2022ydq}. The primary and secondary masses are for the source frame. The SNR is given for each detector individually (H for LIGO Hanford, L for LIGO Livingston, and V for Virgo). The Network SNR is their quadrature sum. The length of data used is noted.}
\begin{ruledtabular}
\begin{tabular}{ccccccccc}
 Time (UTC)& $m_{1}$ ($\textup{M}_\odot$)& $m_{2}$ ($\textup{M}_\odot$)&$D_L$ (Mpc)&H SNR
&L SNR&V SNR&Network SNR& Data length (s) \\  \hline
2023-11-09 23:54:56.048&$1.76^{+0.37}_{-0.22}$&$1.19^{+0.16}_{-0.18}$&$178.08^{+43.59}_{-43.65}$&6.40 &6.14& & 8.87&128 \\
2023-08-14 00:43:53.584&$0.89^{+0.14}_{-0.12}$&$0.15^{+0.02}_{-0.01}$&$65.57^{+18.10}_{-17.79}$&6.96&5.49& & 8.70&512 \\
2020-03-08 18:05:53.02&$0.47^{+0.09}_{-0.05}$&$0.34^{+0.03}_{-0.05}$&$90.83^{+23.67}_{-27.32}$&5.73&5.62&2.14& 8.19&256
\end{tabular}
\end{ruledtabular}
\end{table*}

\begin{table*}
\caption{\label{tab:mc_and_chi_others} Presented here are the comparisons of the spins and detector frame chirp masses as given by the triggers, at 2023-11-09 23:54:56.048 UTC~\cite{Niu:2025nha}, 2023-08-14 00:43:53.584 UTC~\cite{Kacanja:2026byy}, and 2020-03-08 18:05:53.02~\cite{LVK:2022ydq}, as well as our parameter estimation (PE) results.  The effective spin is estimated for the triggers from Eq.~\ref{eq:chi}. The "trigger" values for 2023-11-09 23:54:56.048 UTC correspond to the parameter estimation results, using low-spin priors, as presented in~\cite{Niu:2025nha}. }
\begin{ruledtabular}
\begin{tabular}{ccccc}
 Time (UTC)& $m_{c}$ ($\textup{M}_\odot$) trigger& $m_{c}$ ($\textup{M}_\odot$) PE&$\chi_{\textup{eff}}$ trigger&$\chi_{\textup{eff}}$ PE \\  \hline
2023-11-09 23:54:56.048&$1.3063^{+0.0003}_{-0.0003}$&$1.30647^{+0.00036}_{-0.0.00029}$&$0.014^{+0.016}_{-0.018}$&$0.03905^{+0.05705}_{-0.0274}$ \\
2023-08-14 00:43:53.584&0.30&$0.29820^{+0.00010}_{-0.00008}$& - &$0.027^{+0.07}_{-0.08}$ \\
2020-03-08 18:05:53.02&0.355&$0.35271^{+0.00004}_{-0.00003}$& 0.46 &$0.39^{+0.03}_{-0.02}$
\end{tabular}
\end{ruledtabular}
\end{table*}

\subsubsection{Analysis of single detector data}
For the events 2023-11-09 23:54:56.048 UTC~\cite{Niu:2025nha} and 2023-08-14 00:43:53.584 UTC~\cite{Kacanja:2026byy} we also conduct parameter estimation using only single detector data, H1 and L1. We also attempted this study with the trigger at 2020-03-08 18:05:53.02 UTC, but we were unable to generate decent parameter estimation results.

\begin{figure}[tb]
        \centering
    	\includegraphics[width=\columnwidth]{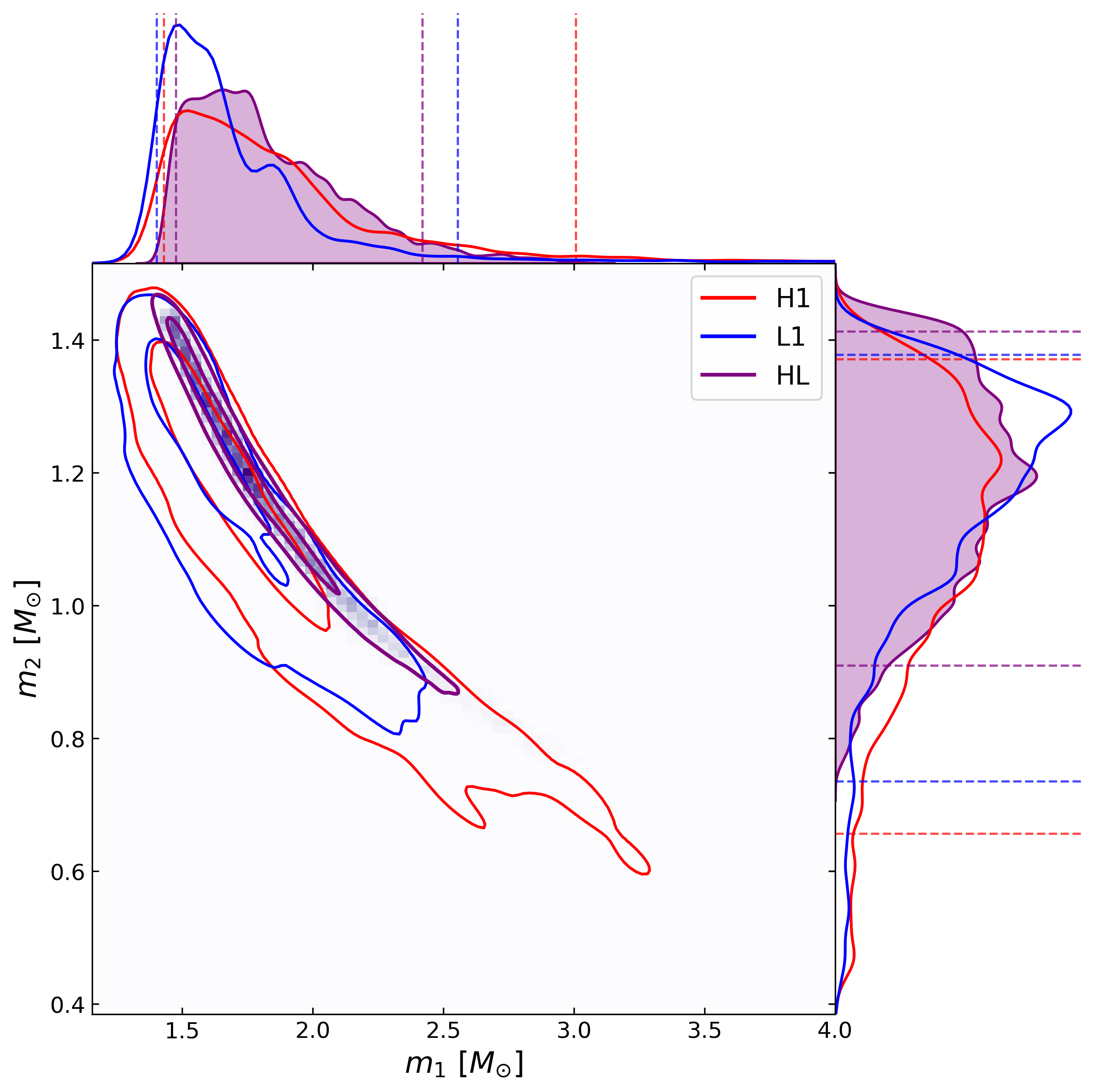}
    	\caption{Posterior distributions for the detector frame component masses, $m_1$ and $m_2$, for the trigger at 2023-11-09 23:54:56.048 UTC. The estimates come from analyzing only the data from L1, only from H1, and analyzing H1 and L1 together.}
        \label{fig:13836093148_128_1s}
    \end{figure}

The results for 2023-11-09 23:54:56.048 UTC~\cite{Niu:2025nha} are presented in Table~\ref{tab:LIGO_1D_ex2} and Fig.~\ref{fig:13836093148_128_1s}. The prior range for the chirp mass is set as $1.2 ~ \textup{M}_\odot < m_c < 1.4 ~ \textup{M}_\odot $. The single detector results (excluding the luminosity distance) are very similar to the two-detector results presented in Tables~\ref{tab:table_PE_others} and \ref{tab:mc_and_chi_others}. This event, if real, would be a binary neutron star. However, the smaller compact object does have some probability to be close to $1.0 ~ \textup{M}_\odot$.

\begin{table*}
\caption{\label{tab:LIGO_1D_ex2} Presented are the parameter estimation results for the LIGO trigger at 2023-11-09 23:54:56.048 UTC~\cite{Niu:2025nha}. The data from only one detector are analyzed. The chirp mass $m_c$ is for the detector frame, while the component masses $m_1$ and $m_2$ are from the source frame.}
\begin{ruledtabular}
\begin{tabular}{cccccccc}
 Detector& $m_{c}$ ($\textup{M}_\odot$)& $m_{1}$ ($\textup{M}_\odot$)&$m_{2}$ ($\textup{M}_\odot$)&$\chi_{\textup{eff}}$&$D_L$ (Mpc)&SNR&Data length (s) \\  \hline
H1&$1.31^{+0.008}_{-0.0009}$&$1.76^{+0.50}_{-0.26}$&$1.12^{+0.18}_{-0.24}$&$0.03^{+0.08}_{-0.04}$&$346.42^{+197.59}_{-124.39}$ & 6.16&128\\
L1&$1.31^{+0.0073}_{-0.0065}$&$1.61^{+0.32}_{-0.16}$&$1.20^{+0.12}_{-0.20}$&$0.05^{+0.04}_{-0.05}$&$374.85^{+214.29}_{-107.73}$ &5.98 & 128
\end{tabular}
\end{ruledtabular}
\end{table*}

\begin{figure}[tb]
        \centering
    	\includegraphics[width=\columnwidth]{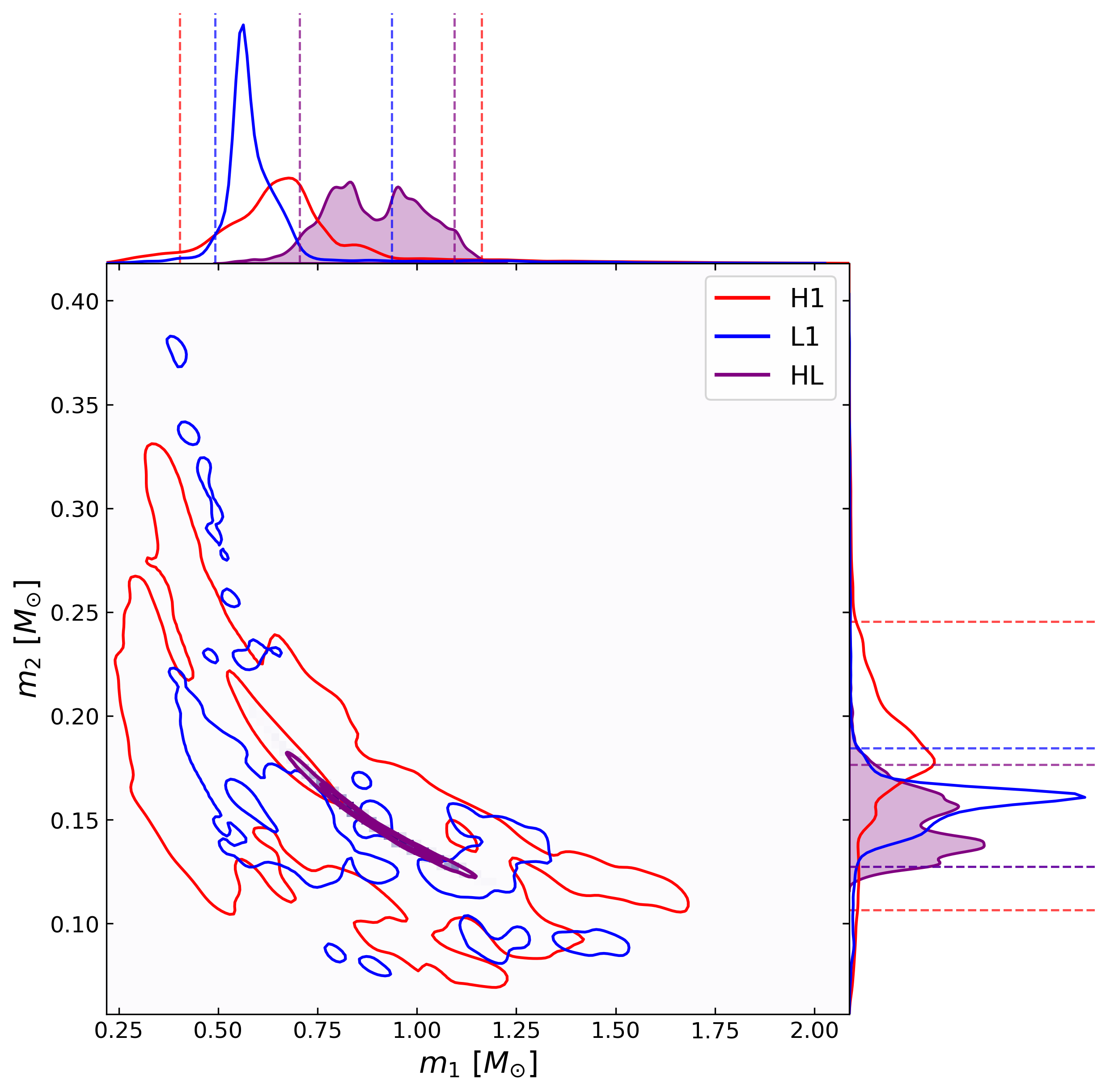}
    	\caption{Posterior distributions for the detector frame component masses, $m_1$ and $m_2$, for the trigger at 2023-08-14 00:43:53.584 UTC. The estimates come from analyzing only the data from L1, only from H1, and analyzing H1 and L1 together.}
        \label{fig:1376009051_512_1s}
    \end{figure}

The single detector parameter estimation results for the trigger 2023-08-14 00:43:53.584 UTC~\cite{Kacanja:2026byy} are presented in Table~\ref{tab:LIGO_1D_ex4} and Fig.~\ref{fig:1376009051_512_1s}. There is agreement between the parameter estimation results using the combined H1-L1 data, and those using the data of H1 and L1 alone. The results from both the H1 data and the L1 data show some bimodality. The primary compact object mass estimate is lower when analyzing the H1 and L1 data alone. For the L1 results, the detector frame chirp mass estimate of $m_c = 0.26 ~ \textup{M}_\odot$ is noticeably low. If this event is real, both compact objects are sub-solar mass.

\begin{table*}
\caption{\label{tab:LIGO_1D_ex4} Presented are the parameter estimation results for the LIGO trigger at 2023-08-14 00:43:53.584 UTC~\cite{Kacanja:2026byy}. The data from only one detector are analyzed. The chirp mass $m_c$ is for the detector frame, while the component masses $m_1$ and $m_2$ are from the source frame. For the analysis of the H1 and L1 data the chirp mass prior was $0.2 ~ \textup{M}_\odot < m_c < 0.4 ~ \textup{M}_\odot$. }
\begin{ruledtabular}
\begin{tabular}{cccccccc}
 Detector& $m_{c}$ ($\textup{M}_\odot$)& $m_{1}$ ($\textup{M}_\odot$)&$m_{2}$ ($\textup{M}_\odot$)&$\chi_{\textup{eff}}$&$D_L$ (Mpc)&SNR&Data length (s) \\  \hline
H1&$0.295^{+0.003}_{-0.003}$&$0.66^{+0.15}_{-0.14}$&$0.18^{+0.03}_{-0.03}$&-$0.14^{+0.02}_{-0.01}$&$98.28^{+636.94}_{-37.57}$ & 6.33&512\\
L1&$0.26^{+0.003}_{-0.003}$&$0.57^{+0.09}_{-0.03}$&$0.16^{+0.01}_{-0.02}$&$-0.14^{+0.08}_{-0.04}$&$96.81^{+63.42}_{-26.77}$ &6.16 & 512
\end{tabular}
\end{ruledtabular}
\end{table*}

\section{Supplemental Figures}\label{appendix}
\label{App:SuppFigs}

Presented here are the remaining parameter estimation plots from the analyses above.

\begin{figure}[tb]
        \centering
    	\includegraphics[width=\columnwidth]{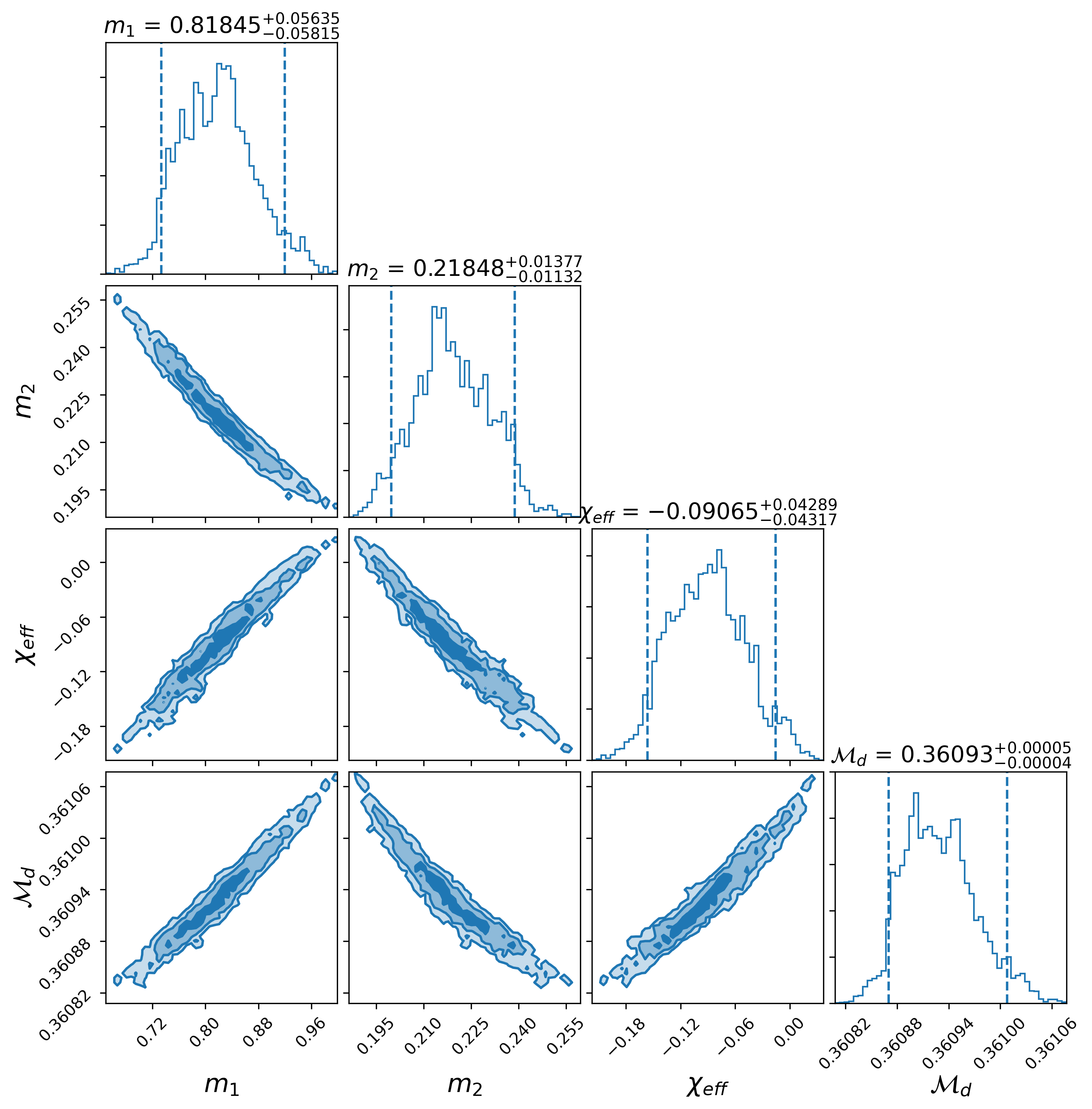}
    	\caption{A corner plot for the trigger at 2023-07-19 11:50:50.27 UTC showing the source frame component masses, $m_1$ and $m_2$, the effective spin $\chi_{\rm eff}$, and the detector frame chirp mass $m_c$.}
        \label{fig:1373802668_256_2}
    \end{figure}

\begin{figure}[tb]
        \centering
    	\includegraphics[width=\columnwidth]{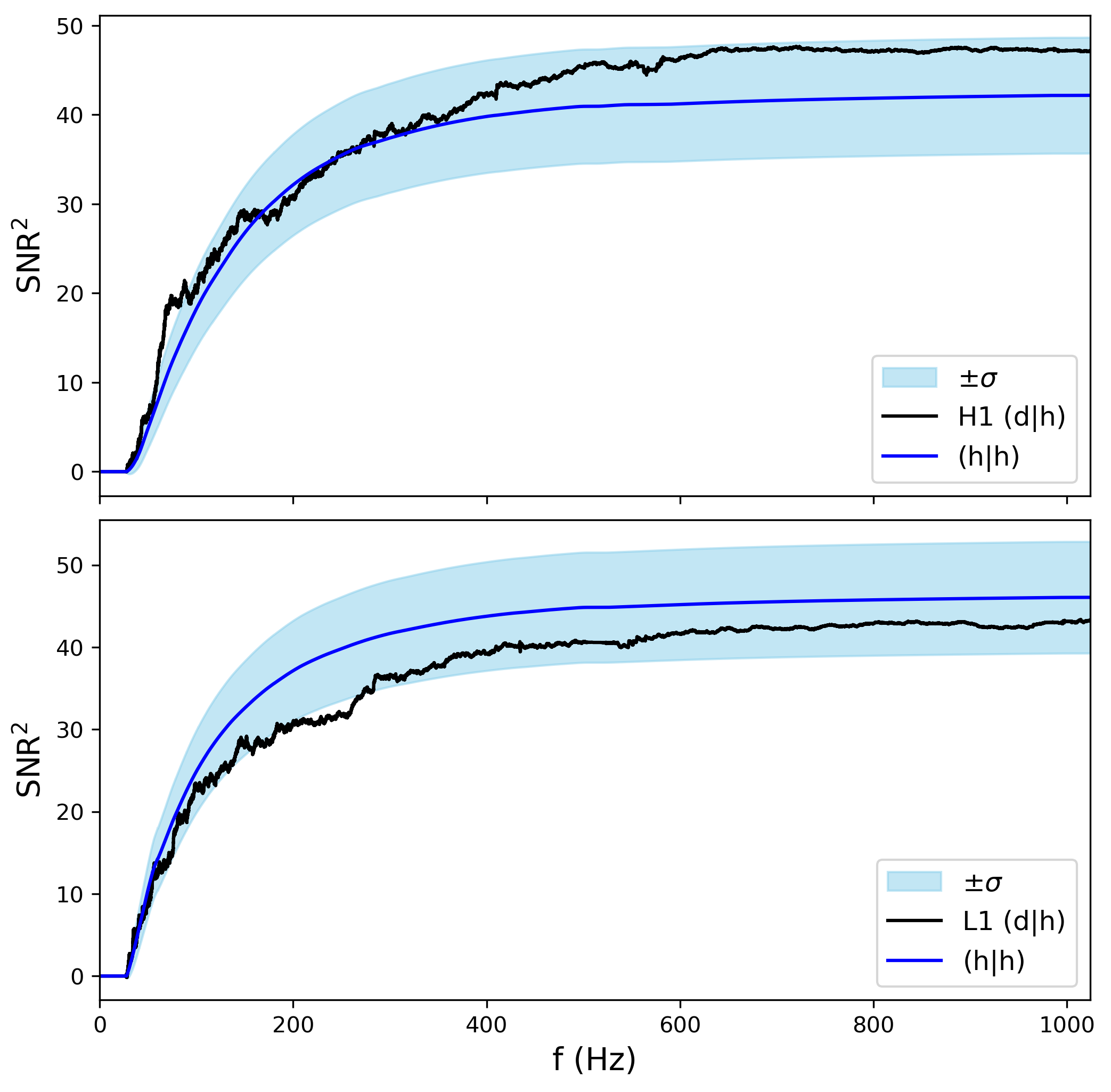}
    	\caption{The SNR$^2$ of the signal as a function of frequency for the  2023-07-19 11:50:50.27 UTC trigger. The black line is calculated from the data ($d \mid h$). The blue line corresponds to a draw from the likelihood ($h \mid h$). The blue region are the $\pm 1 \sigma$ errors on the SNR$^2$ estimation.}
        \label{fig:1373802668_SNR2_H1L1_together}
    \end{figure}

\begin{figure}[tb]
        \centering
    	\includegraphics[width=\columnwidth]{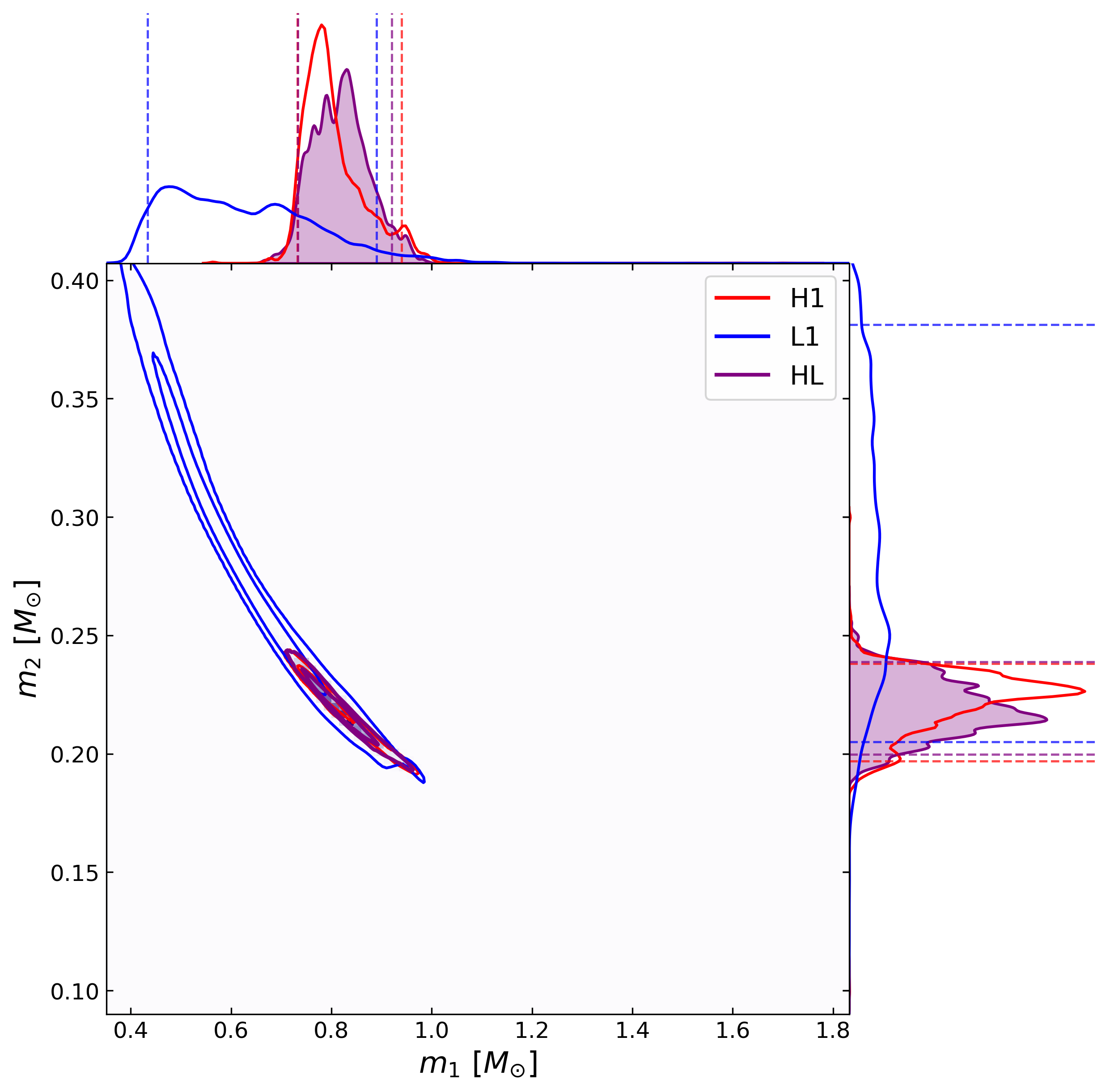}
    	\caption{Posterior distributions for the detector frame component masses, $m_1$ and $m_2$, for the trigger at 2023-07-19 11:50:50.27 UTC. The estimates come from analyzing only the data from L1, only from H1, and analyzing H1 and L1 together.}
        \label{fig:1373802668_256_1s}
    \end{figure}
    
    \begin{figure}[tb]
        \centering
    	\includegraphics[width=\columnwidth]{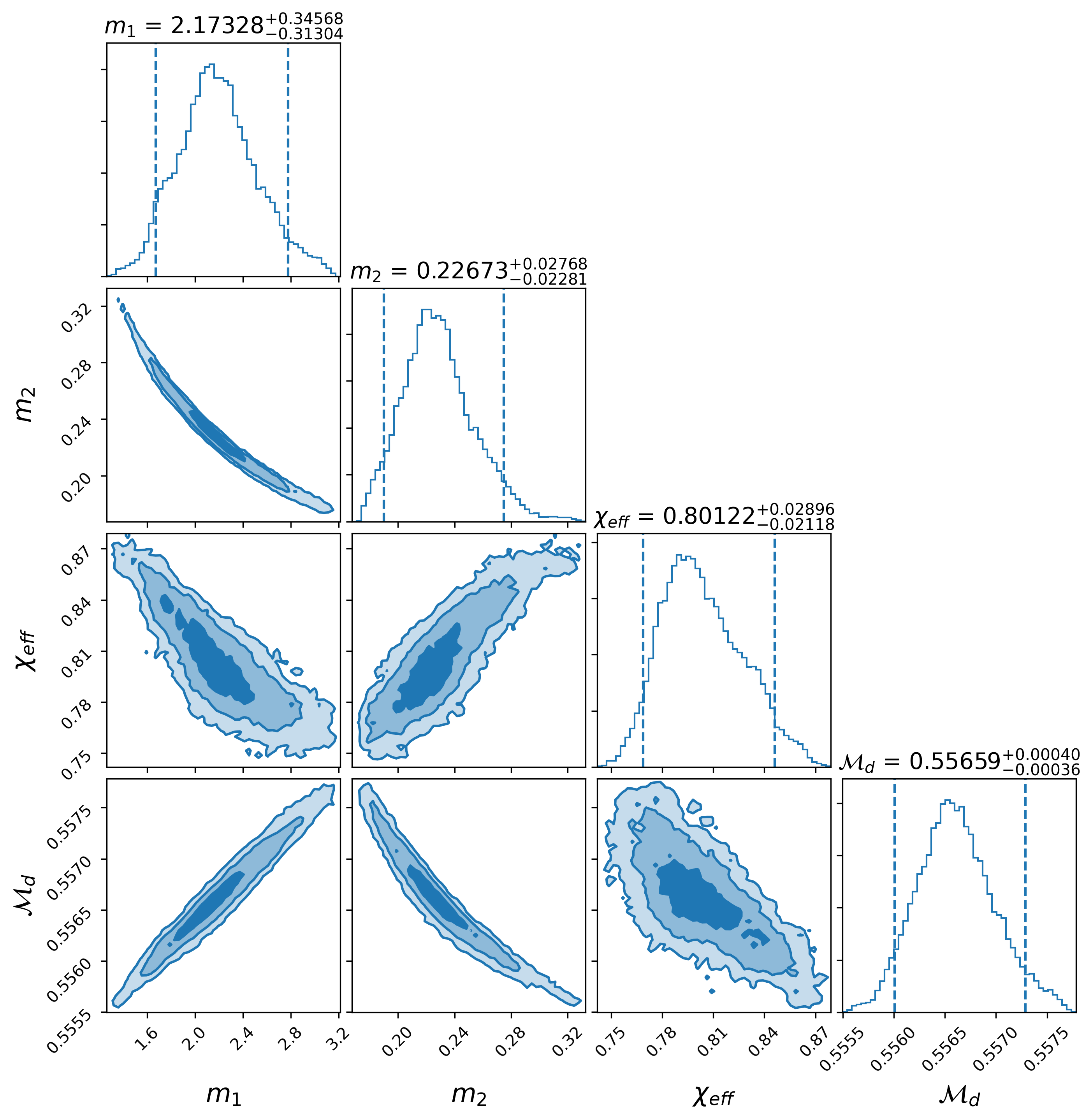}
    	\caption{A corner plot for the trigger at 2023-10-14 08:15:06.33 UTC showing the source frame component masses, $m_1$ and $m_2$, the effective spin $\chi_{\rm eff}$, and the detector frame chirp mass $m_c$.}
        \label{fig:1381306524_256_2}
    \end{figure}
    
    \begin{figure}[tb]
        \centering
    	\includegraphics[width=\columnwidth]{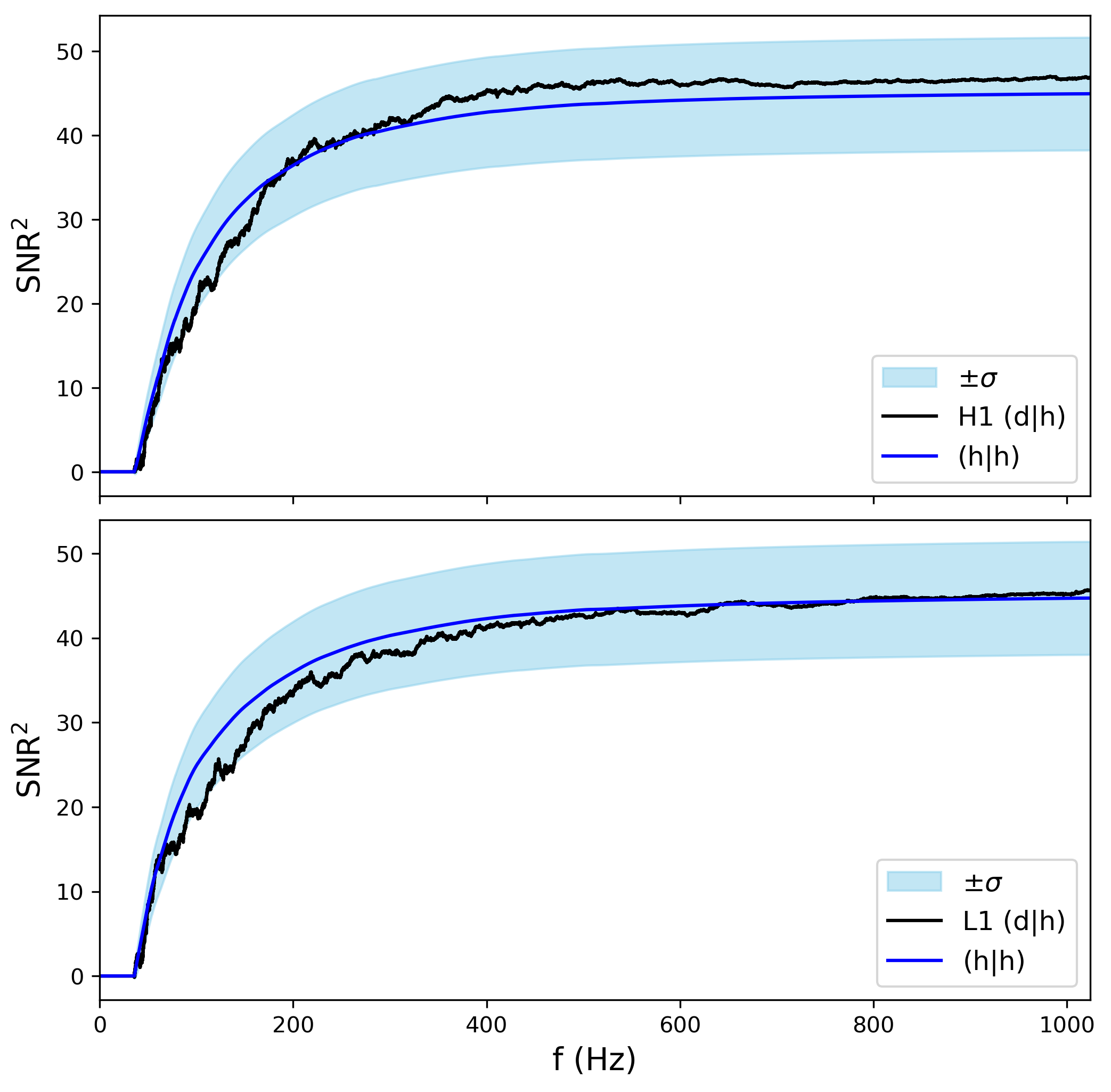}
    	\caption{The SNR$^2$ of the signal as a function of frequency for the  2023-10-14 08:15:06.33 UTC trigger. The black line is calculated from the data ($d \mid h$). The blue line corresponds to a draw from the likelihood ($h \mid h$). The blue region are the $\pm 1 \sigma$ errors on the SNR$^2$ estimation.}
        \label{fig:1381306524_SNR2_H1L1_together}
    \end{figure}
    
    \begin{figure}[tb]
        \centering
    	\includegraphics[width=\columnwidth]{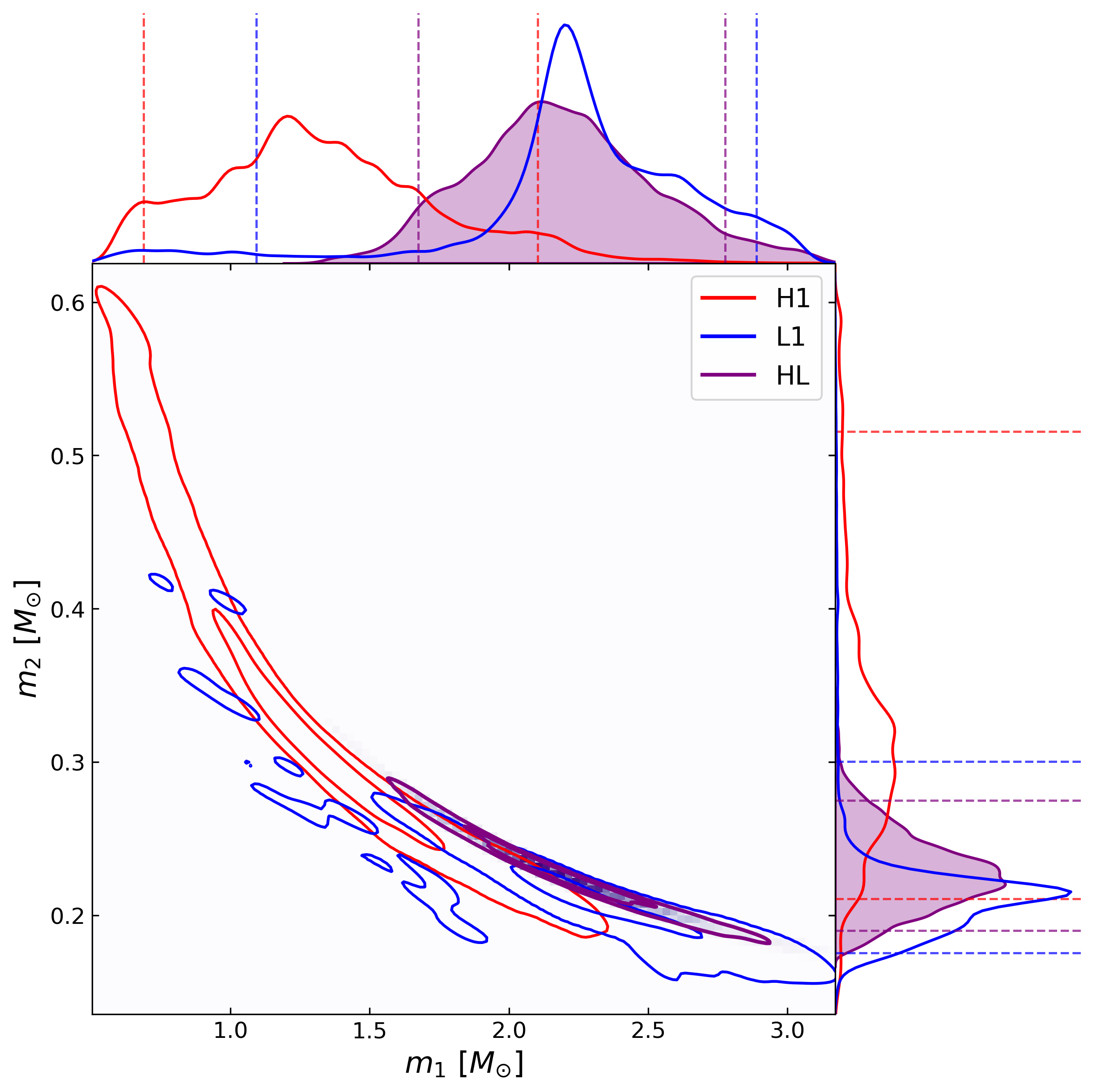}
    	\caption{Posterior distributions for the detector frame component masses, $m_1$ and $m_2$, for the trigger at 2023-10-14 08:15:06.33 UTC. The estimates come from analyzing only the data from L1, only from H1, and analyzing H1 and L1 together.}
        \label{fig:1381306524_256_1s}
    \end{figure}
    
    \begin{figure}[tb]
        \centering
    	\includegraphics[width=\columnwidth]{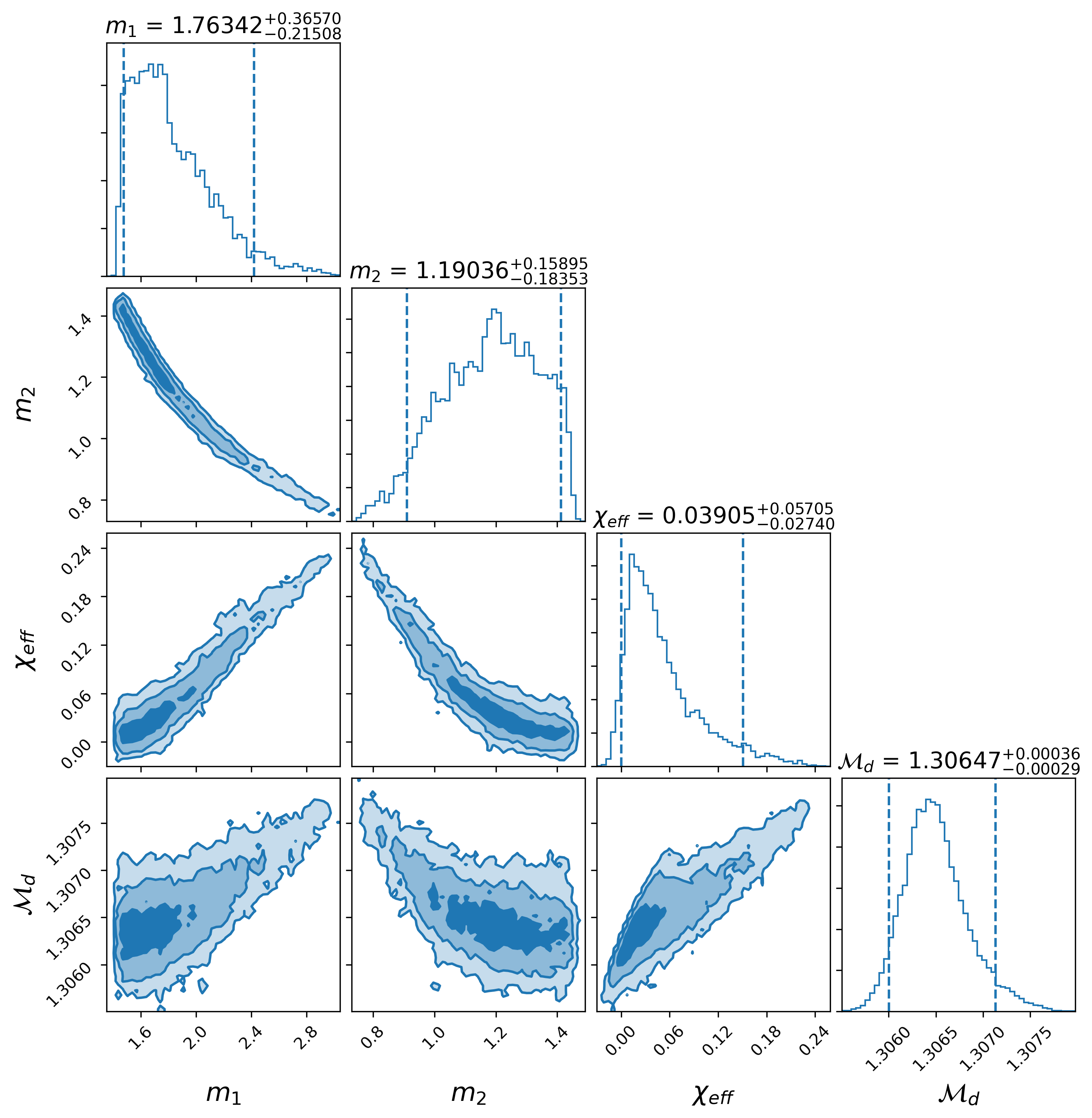}
    	\caption{A corner plot for the trigger at 2023-11-09 23:54:56.048 UTC showing the source frame component masses, $m_1$ and $m_2$, the effective spin $\chi_{\rm eff}$, and the detector frame chirp mass $m_c$.}
        \label{fig:1383609314_128_2}
    \end{figure}
    
    \begin{figure}[tb]
        \centering
    	\includegraphics[width=\columnwidth]{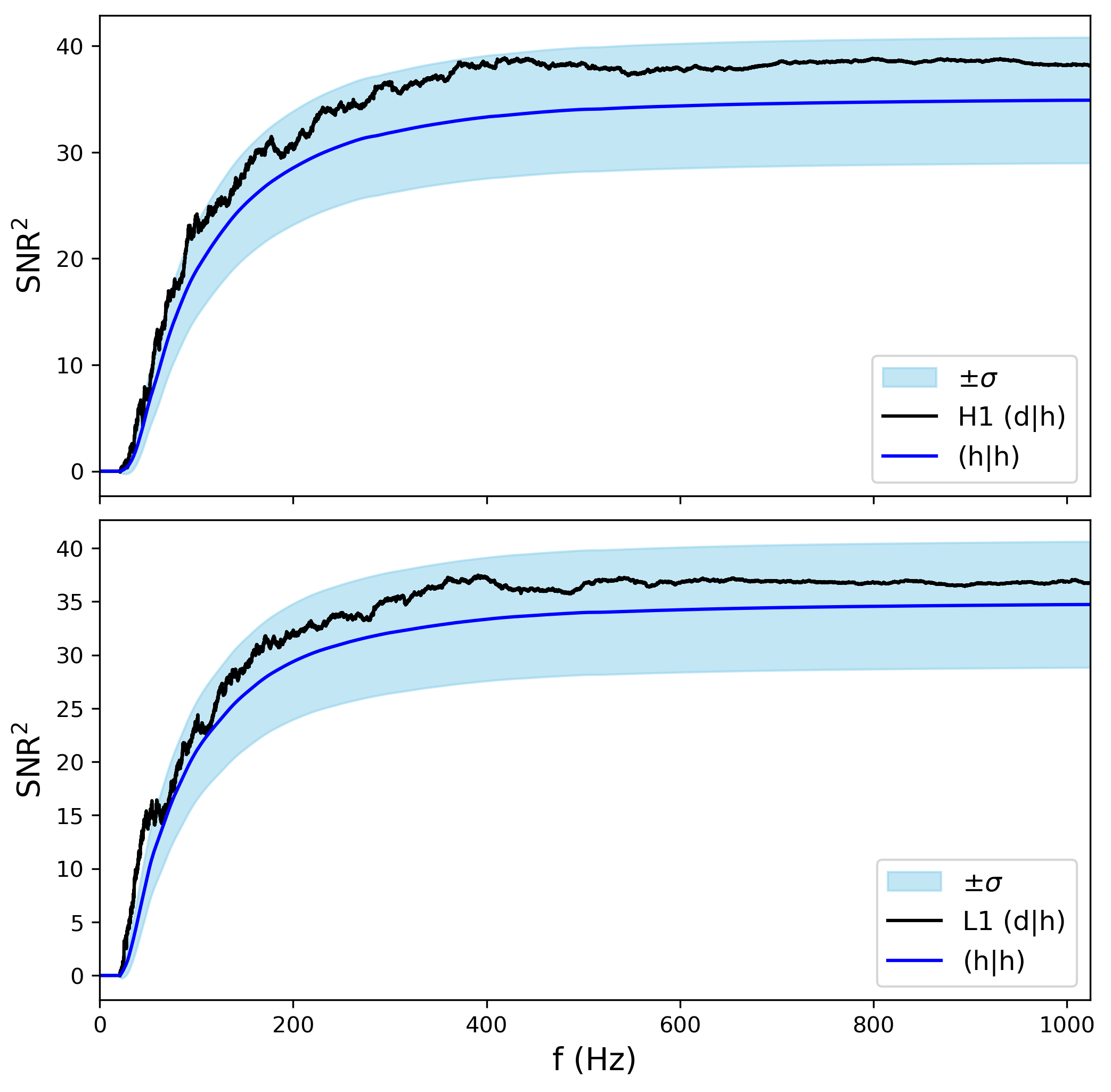}
    	\caption{The SNR$^2$ of the signal as a function of frequency for the  2023-11-09 23:54:56.048 UTC trigger. The black line is calculated from the data ($d \mid h$). The blue line corresponds to a draw from the likelihood ($h \mid h$). The blue region are the $\pm 1 \sigma$ errors on the SNR$^2$ estimation.}
        \label{fig:1383609314_SNR2_H1L1_together}
    \end{figure}
    
    \begin{figure}[tb]
        \centering
    	\includegraphics[width=\columnwidth]{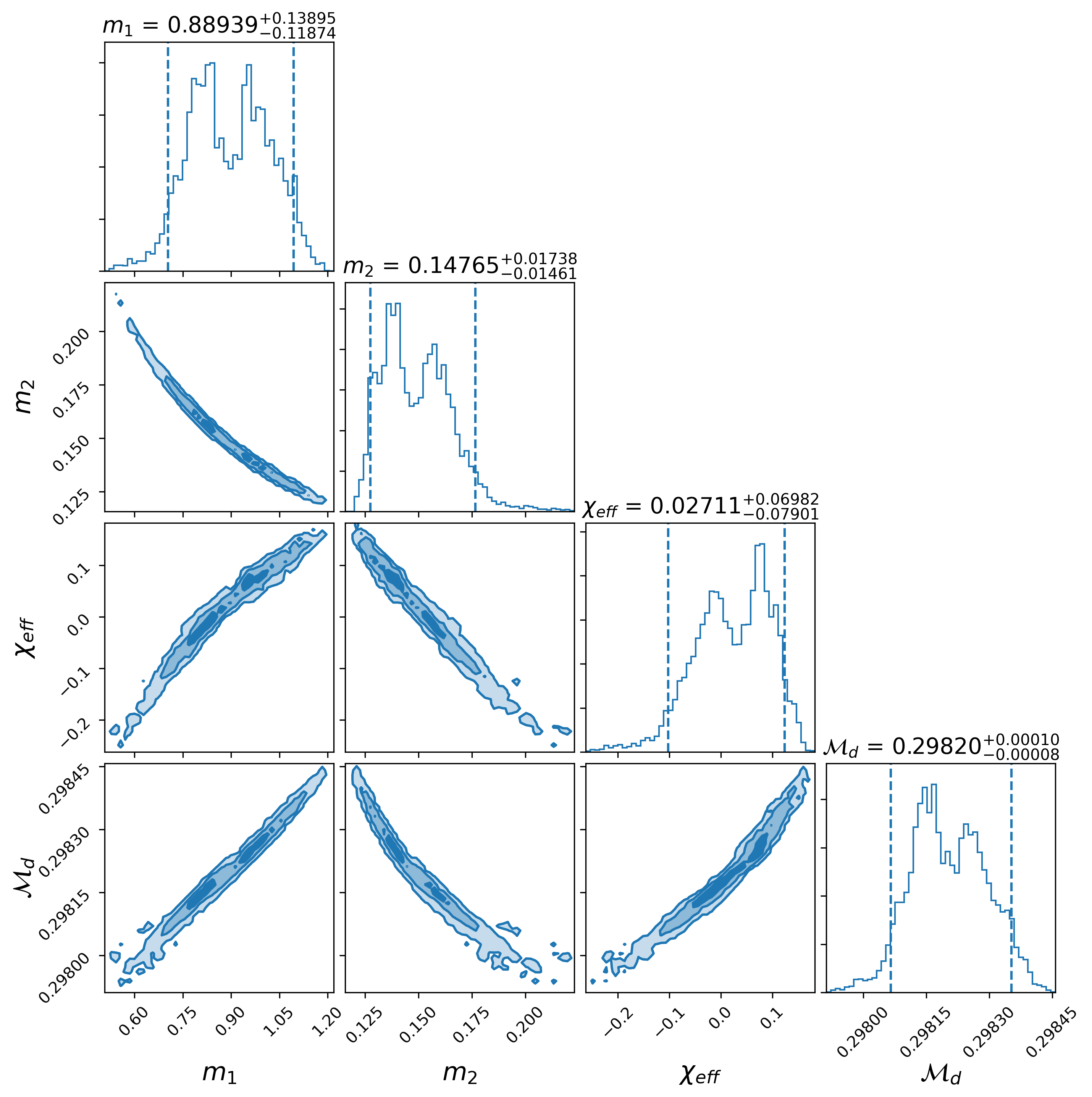}
    	\caption{A corner plot for the trigger at 2023-08-14 00:43:53.584 UTC showing the source frame component masses, $m_1$ and $m_2$, the effective spin $\chi_{\rm eff}$, and the detector frame chirp mass $m_c$.}
        \label{fig:1376009051_512_2}
    \end{figure}
    
     \begin{figure}[tb]
        \centering
    	\includegraphics[width=\columnwidth]{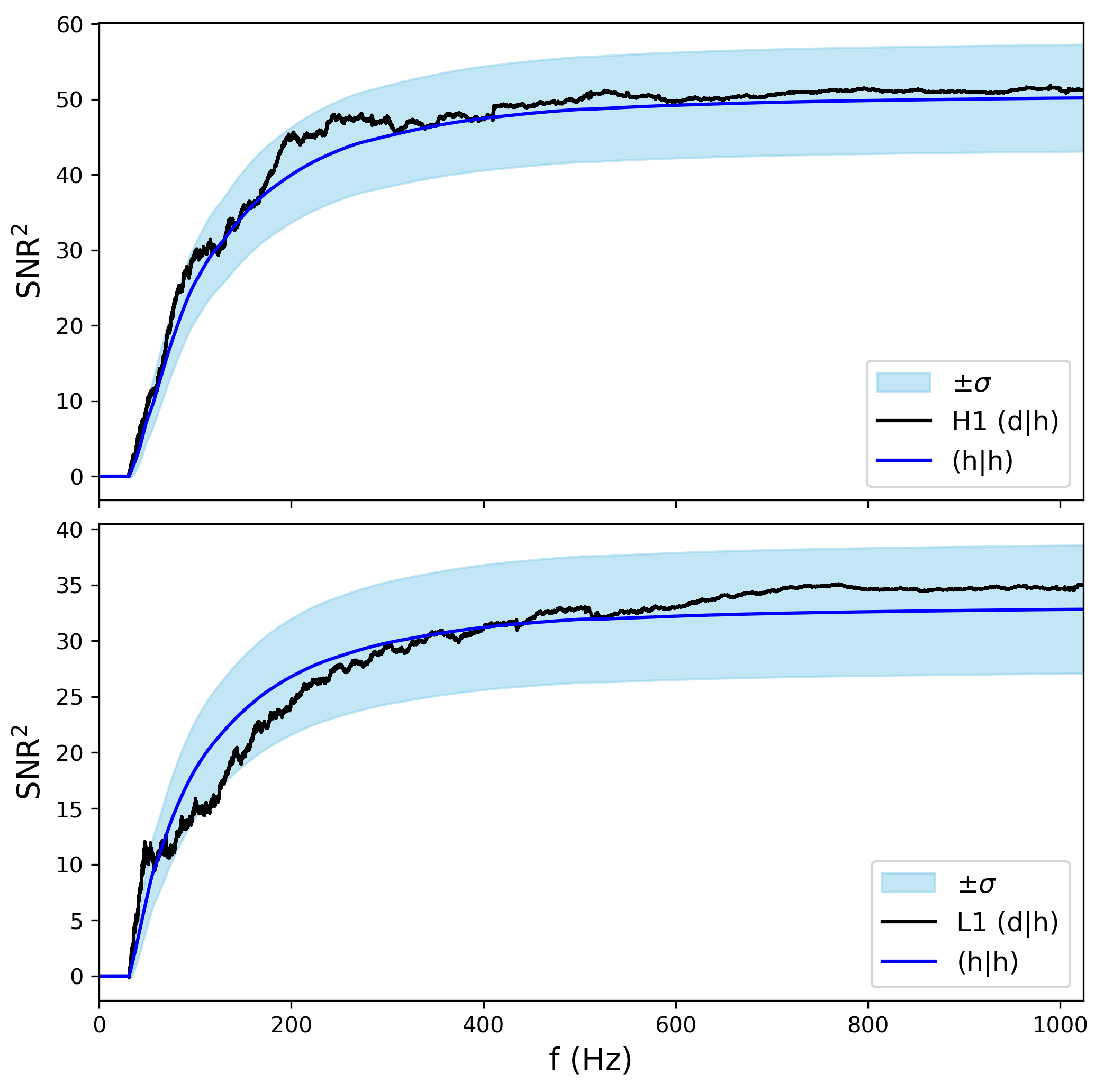}
    	\caption{The SNR$^2$ of the signal as a function of frequency for the  2023-08-14 00:43:53.584 UTC trigger. The black line is calculated from the data ($d \mid h$). The blue line corresponds to a draw from the likelihood ($h \mid h$). The blue region are the $\pm 1 \sigma$ errors on the SNR$^2$ estimation.}
        \label{fig:1376009051_SNR2_H1L1_together}
    \end{figure}
    
     \begin{figure}[tb]
        \centering
    	\includegraphics[width=\columnwidth]{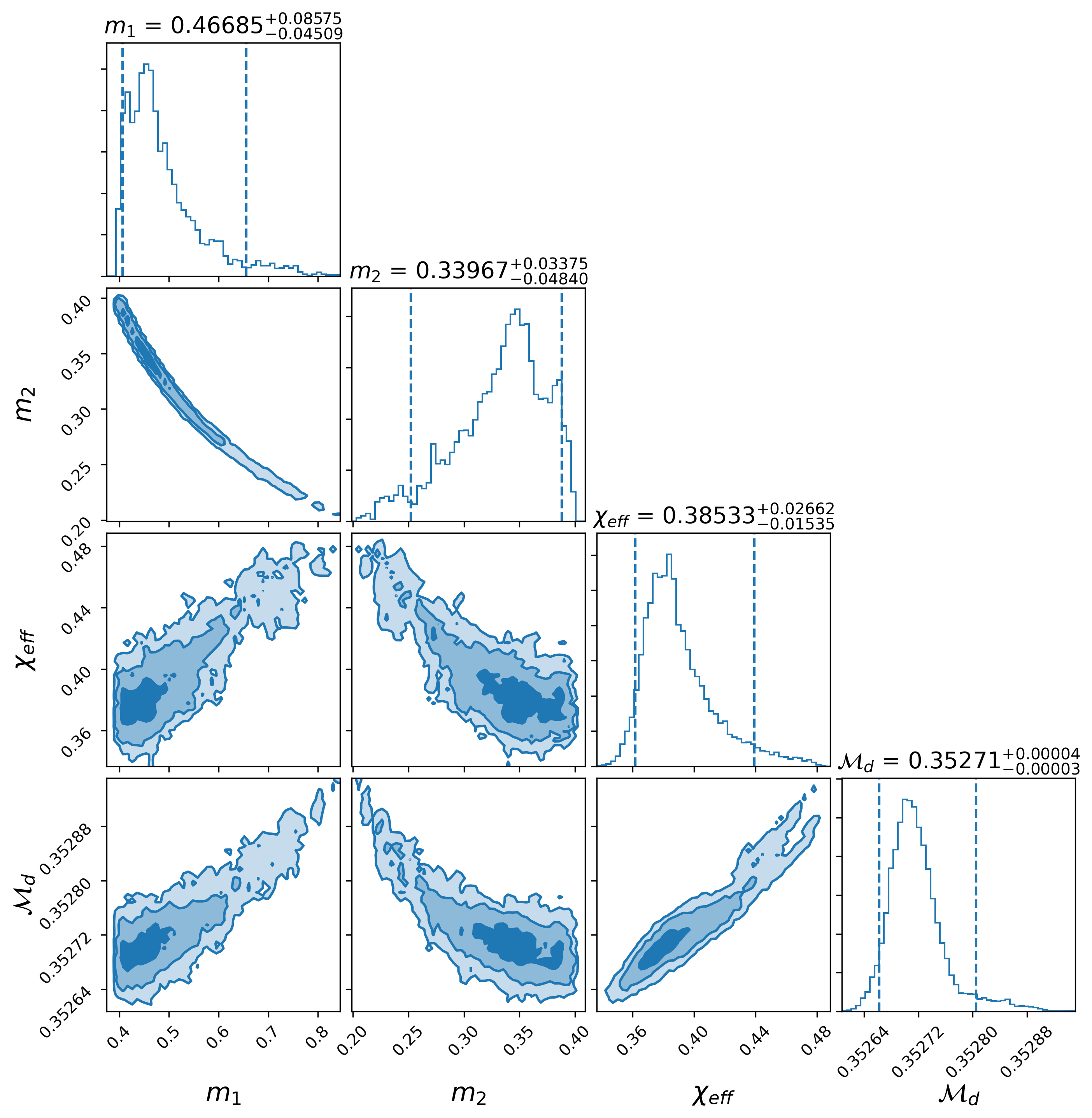}
    	\caption{A corner plot for the trigger at 2020-03-08 18:05:53.02 UTC showing the source frame component masses, $m_1$ and $m_2$, the effective spin $\chi_{\rm eff}$, and the detector frame chirp mass $m_c$.}
        \label{fig:1267725971_256_2}
    \end{figure}
    
         \begin{figure}[tb]
        \centering
    	\includegraphics[width=\columnwidth]{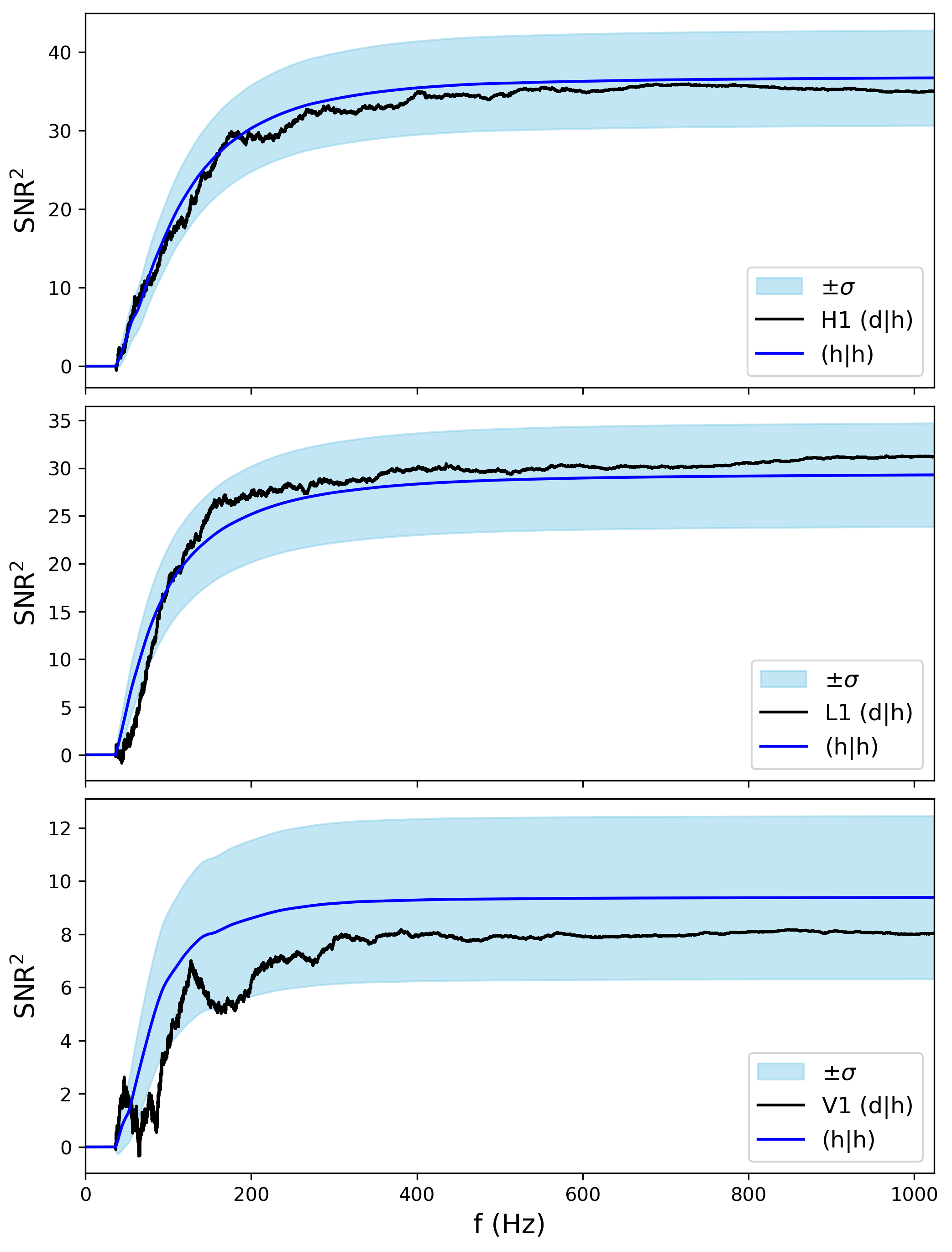}
    	\caption{The SNR$^2$ of the signal as a function of frequency for the 2020-03-08 18:05:53.02 UTC trigger. The black line is calculated from the data ($d \mid h$). The blue line corresponds to a draw from the likelihood ($h \mid h$). The blue region are the $\pm 1 \sigma$ errors on the SNR$^2$ estimation.}
        \label{fig:1267725971_SNR2_H1L1V1_together}
    \end{figure}

\end{appendix}

\bibliography{apssamp}

@PREAMBLE{
 "\providecommand{\noopsort}[1]{}" 
 # "\providecommand{\singleletter}[1]{#1}%" 
}

@article{Kashlinsky:2005di,
    author = "Kashlinsky, Alexander and Arendt, R. G. and Mather, John C. and Moseley, S. H.",
    title = "{Tracing the first stars with fluctuations of the cosmic infrared background}",
    eprint = "astro-ph/0511105",
    archivePrefix = "arXiv",
    doi = "10.1038/nature04143",
    journal = "Nature",
    volume = "438",
    pages = "45--50",
    year = "2005"
}

@article{Kashlinsky:2016sdv,
    author = "Kashlinsky, A.",
    title = "{LIGO gravitational wave detection, primordial black holes and the near-IR cosmic infrared background anisotropies}",
    eprint = "1605.04023",
    archivePrefix = "arXiv",
    primaryClass = "astro-ph.CO",
    doi = "10.3847/2041-8205/823/2/L25",
    journal = "Astrophys. J. Lett.",
    volume = "823",
    number = "2",
    pages = "L25",
    year = "2016"
}

@article{Hawking:1971ei,
    author = "Hawking, Stephen",
    title = "{Gravitationally collapsed objects of very low mass}",
    doi = "10.1093/mnras/152.1.75",
    journal = "Mon. Not. Roy. Astron. Soc.",
    volume = "152",
    pages = "75",
    year = "1971"
}

@article{Carr:1974nx,
  author        = {Carr, B. J. and Hawking, S. W.},
  title         = "{Black Holes in the Early Universe}",
  journal       = {Mon. Not. Roy. Astron. Soc.},
  volume        = {168},
  pages         = {399--415},
  year          = {1974},
  doi           = {10.1093/mnras/168.2.399}
}

@article{Carr:2016drx,
  author        = {Carr, Bernard and K{\"u}hnel, Florian and Sandstad, Marit},
  title         = "{Primordial Black Holes as Dark Matter}",
  journal       = {Phys. Rev. D},
  volume        = {94},
  number        = {8},
  pages         = {083504},
  year          = {2016},
  doi           = {10.1103/PhysRevD.94.083504},
  eprint        = {1607.06077},
  archivePrefix = {arXiv},
  primaryClass  = {astro-ph.CO}
}

@article{Carr:2020xqk,
  author        = {Carr, Bernard and K{\"u}hnel, Florian},
  title         = "{Primordial Black Holes as Dark Matter: Recent Developments}",
  journal       = {Ann. Rev. Nucl. Part. Sci.},
  volume        = {70},
  pages         = {355--394},
  year          = {2020},
  doi           = {10.1146/annurev-nucl-050520-125911},
  eprint        = {2006.02838},
  archivePrefix = {arXiv},
  primaryClass  = {astro-ph.CO}
}

@incollection{Escriva:2022duf,
  author        = {Escriv{\`a}, Albert and K{\"u}hnel, Florian and Tada, Yuichiro},
  title         = "{Primordial Black Holes}",
  booktitle     = {Black Holes in the Era of Gravitational-Wave Astronomy},
  publisher     = {Elsevier},
  pages         = {261--377},
  year          = {2024},
  doi           = {10.1016/B978-0-32-395636-9.00012-8},
  eprint        = {2211.05767},
  archivePrefix = {arXiv},
  primaryClass  = {astro-ph.CO}
}

@article{Carr:2023tpt,
  author        = {Carr, Bernard and Clesse, Sebastien and Garc{\'i}a-Bellido, Juan and Hawkins, Michael R. S. and K{\"u}hnel, Florian},
  title         = "{Observational Evidence for Primordial Black Holes: A Positivist Perspective}",
  journal       = {Phys. Rept.},
  volume        = {1054},
  pages         = {1--68},
  year          = {2024},
  doi           = {10.1016/j.physrep.2023.11.005},
  eprint        = {2306.03903},
  archivePrefix = {arXiv},
  primaryClass  = {astro-ph.CO}
}

@article{2017ApJ...836L..18M,
  author        = {Mediavilla, E. and Jim{\'e}nez-Vicente, J. and Mu{\~n}oz, J. A. and Vives-Arias, H. and Calder{\'o}n-Infante, J.},
  title         = "{Limits on the Mass and Abundance of Primordial Black Holes from Quasar Gravitational Microlensing}",
  journal       = {Astrophys. J. Lett.},
  volume        = {836},
  number        = {2},
  pages         = {L18},
  year          = {2017},
  doi           = {10.3847/2041-8213/aa5dab},
  eprint        = {1702.00947},
  archivePrefix = {arXiv},
  primaryClass  = {astro-ph.CO}
}

@article{Cappelluti:2013bda,
  author        = {Cappelluti, N. and Kashlinsky, A. and Arendt, R. G. and Comastri, A. and Fazio, G. G. and Finoguenov, A. and Hasinger, G. and Mather, J. C. and Miyaji, T. and Moseley, S. H.},
  title         = "{Cross-correlating Cosmic IR and X-ray Background Fluctuations: Evidence of Significant Black Hole Populations among the CIB Sources}",
  journal       = {Astrophys. J.},
  volume        = {769},
  pages         = {68},
  year          = {2013},
  doi           = {10.1088/0004-637X/769/1/68},
  eprint        = {1210.5302},
  archivePrefix = {arXiv},
  primaryClass  = {astro-ph.CO}
}

@article{Hasinger:2020ptw,
  author        = {Hasinger, G.},
  title         = "{Illuminating the Dark Ages: Cosmic Backgrounds from Accretion onto Primordial Black Hole Dark Matter}",
  journal       = {JCAP},
  volume        = {07},
  pages         = {022},
  year          = {2020},
  doi           = {10.1088/1475-7516/2020/07/022},
  eprint        = {2003.05150},
  archivePrefix = {arXiv},
  primaryClass  = {astro-ph.CO}
}

@article{2024Natur.633..318C,
  author        = {Carniani, Stefano and Hainline, Kevin and D'Eugenio, Francesco and Eisenstein, Daniel and Jakobsen, Peter and others},
  title         = "{Spectroscopic Confirmation of Two Luminous Galaxies at a Redshift of 14}",
  journal       = {Nature},
  volume        = {633},
  pages         = {318--322},
  year          = {2024},
  doi           = {10.1038/s41586-024-07860-9},
  eprint        = {2405.18485},
  archivePrefix = {arXiv},
  primaryClass  = {astro-ph.GA}
}

@article{2024Natur.628...57F,
  author        = {Furtak, Lukas J. and Labb{\'e}, Ivo and Zitrin, Adi and Greene, Jenny E. and Dayal, Pratika and Chemerynska, Iryna and Kokorev, Vasily and Miller, Tim B. and Goulding, Andy D. and de Graaff, Anna and others},
  title         = "{A High Black-Hole-to-Host Mass Ratio in a Lensed AGN in the Early Universe}",
  journal       = {Nature},
  volume        = {628},
  number        = {8006},
  pages         = {57--61},
  year          = {2024},
  doi           = {10.1038/s41586-024-07184-8},
  eprint        = {2308.05735},
  archivePrefix = {arXiv},
  primaryClass  = {astro-ph.GA}
}

@article{Juodzbalis:2025qso1,
   title={A direct black-hole mass measurement in a little red dot at high redshift},
   volume={653},
   ISSN={1476-4687},
   url={http://dx.doi.org/10.1038/s41586-026-10579-4},
   DOI={10.1038/s41586-026-10579-4},
   number={8116},
   journal={Nature},
   publisher={Springer Science and Business Media LLC},
   author={Juodžbalis, Ignas and others},
   year={2026},
   month=May, pages={1017–1021} }

@article{Maiolino:2025qso1,
    author = "Maiolino, Roberto and others",
    title = "{A black hole in a near pristine galaxy 700 Myr after the big bang}",
    eprint = "2505.22567",
    archivePrefix = "arXiv",
    primaryClass = "astro-ph.GA",
    doi = "10.1093/mnras/staf2109",
    journal = "Mon. Not. Roy. Astron. Soc.",
    volume = "548",
    number = "1",
    pages = "staf2109",
    year = "2026"
}

@article{Zhang:2025oyl,
    author = {Zhang, Saiyang and Liu, Boyuan and Bromm, Volker and K{\"u}hnel, Florian},
    title = "{Primordial Black Holes as Seeds for Extremely Overmassive Active Galactic Nuclei Observed by JWST}",
    eprint = "2512.14066",
    archivePrefix = "arXiv",
    primaryClass = "astro-ph.GA",
    doi = "10.3847/2041-8213/ae4bd0",
    journal = "Astrophys. J. Lett.",
    volume = "1000",
    number = "1",
    pages = "L19",
    year = "2026"
}

@article{Carr:2019kxo,
  author        = {Carr, Bernard and Clesse, Sebastien and Garc{\'i}a-Bellido, Juan and K{\"u}hnel, Florian},
  title         = "{Cosmic Conundra Explained by Thermal History and Primordial Black Holes}",
  journal       = {Phys. Dark Univ.},
  volume        = {31},
  pages         = {100755},
  year          = {2021},
  doi           = {10.1016/j.dark.2020.100755},
  eprint        = {1906.08217},
  archivePrefix = {arXiv},
  primaryClass  = {astro-ph.CO}
}

@article{LIGOScientific:2014pky,
    author = "Aasi, J. and others",
    collaboration = "LIGO Scientific",
    title = "{Advanced LIGO}",
    eprint = "1411.4547",
    archivePrefix = "arXiv",
    primaryClass = "gr-qc",
    doi = "10.1088/0264-9381/32/7/074001",
    journal = "Class. Quant. Grav.",
    volume = "32",
    pages = "074001",
    year = "2015"
}

@article{KAGRA:2020tym,
    author = "Akutsu, T. and others",
    collaboration = "KAGRA",
    title = "{Overview of KAGRA: Detector design and construction history}",
    eprint = "2005.05574",
    archivePrefix = "arXiv",
    primaryClass = "physics.ins-det",
    doi = "10.1093/ptep/ptaa125",
    journal = "PTEP",
    volume = "2021",
    number = "5",
    pages = "05A101",
    year = "2021"
}

@article{VIRGO:2014yos,
    author = "Acernese, F. and others",
    collaboration = "VIRGO",
    title = "{Advanced Virgo: a second-generation interferometric gravitational wave detector}",
    eprint = "1408.3978",
    archivePrefix = "arXiv",
    primaryClass = "gr-qc",
    doi = "10.1088/0264-9381/32/2/024001",
    journal = "Class. Quant. Grav.",
    volume = "32",
    number = "2",
    pages = "024001",
    year = "2015"
}

@article{Christensen:2022bxb,
    author = "Christensen, Nelson and Meyer, Renate",
    title = "{Parameter estimation with gravitational waves}",
    eprint = "2204.04449",
    archivePrefix = "arXiv",
    primaryClass = "gr-qc",
    doi = "10.1103/RevModPhys.94.025001",
    journal = "Rev. Mod. Phys.",
    volume = "94",
    number = "2",
    pages = "025001",
    year = "2022"
}

@article{KAGRA:2023pio,
    author = "Abbott, R. and others",
    collaboration = "KAGRA, VIRGO, LIGO Scientific",
    title = "{Open Data from the Third Observing Run of LIGO, Virgo, KAGRA, and GEO}",
    eprint = "2302.03676",
    archivePrefix = "arXiv",
    primaryClass = "gr-qc",
    reportNumber = "LIGO-P2200316",
    doi = "10.3847/1538-4365/acdc9f",
    journal = "Astrophys. J. Suppl.",
    volume = "267",
    number = "2",
    pages = "29",
    year = "2023"
}

@article{Cornish:2021wxy,
    author = "Cornish, Neil J.",
    title = "{Rapid and Robust Parameter Inference for Binary Mergers}",
    eprint = "2101.01188",
    archivePrefix = "arXiv",
    primaryClass = "gr-qc",
    doi = "10.1103/PhysRevD.103.104057",
    journal = "Phys. Rev. D",
    volume = "103",
    number = "10",
    pages = "104057",
    year = "2021"
}

@article{LIGOScientific:2025slb,
    author = "Abac, A. G. and others",
    collaboration = "LIGO Scientific, VIRGO, KAGRA",
    title = "{GWTC-4.0: Updating the Gravitational-Wave Transient Catalog with Observations from the First Part of the Fourth LIGO-Virgo-KAGRA Observing Run}",
    eprint = "2508.18082",
    archivePrefix = "arXiv",
    primaryClass = "gr-qc",
    reportNumber = "LIGO-P2400386",
    doi = "10.3847/2041-8213/ae2c74",
    journal = "Astrophys. J. Lett.",
    volume = "1004",
    number = "2",
    pages = "L22",
    year = "2026"
}

@article{LVK:2022ydq,
    author = "Abbott, R. and others",
    collaboration = "LVK",
    title = "{Search for subsolar-mass black hole binaries in the second part of Advanced LIGO{\textquoteright}s and Advanced Virgo{\textquoteright}s third observing run}",
    eprint = "2212.01477",
    archivePrefix = "arXiv",
    primaryClass = "astro-ph.HE",
    doi = "10.1093/mnras/stad588",
    journal = "Mon. Not. Roy. Astron. Soc.",
    volume = "524",
    number = "4",
    pages = "5984--5992",
    year = "2023",
    note = "[Erratum: Mon.Not.Roy.Astron.Soc. 526, 6234 (2023)]"
}

@article{LIGOScientific:2021job,
    author = "Abbott, R. and others",
    collaboration = "LIGO Scientific, VIRGO, KAGRA",
    title = "{Search for Subsolar-Mass Binaries in the First Half of Advanced LIGO{\textquoteright}s and Advanced Virgo{\textquoteright}s Third Observing Run}",
    eprint = "2109.12197",
    archivePrefix = "arXiv",
    primaryClass = "astro-ph.CO",
    reportNumber = "LIGO-P2100163-v8",
    doi = "10.1103/PhysRevLett.129.061104",
    journal = "Phys. Rev. Lett.",
    volume = "129",
    number = "6",
    pages = "061104",
    year = "2022"
}

@article{LIGOScientific:2019kan,
    author = "Abbott, B. P. and others",
    collaboration = "LIGO Scientific, Virgo",
    title = "{Search for Subsolar Mass Ultracompact Binaries in Advanced LIGO{\textquoteright}s Second Observing Run}",
    eprint = "1904.08976",
    archivePrefix = "arXiv",
    primaryClass = "astro-ph.CO",
    reportNumber = "LIGO-P1900037",
    doi = "10.1103/PhysRevLett.123.161102",
    journal = "Phys. Rev. Lett.",
    volume = "123",
    number = "16",
    pages = "161102",
    year = "2019"
}

@article{LIGOScientific:2018glc,
    author = "Abbott, B. P. and others",
    collaboration = "LIGO Scientific, Virgo",
    title = "{Search for Subsolar-Mass Ultracompact Binaries in Advanced LIGO{\textquoteright}s First Observing Run}",
    eprint = "1808.04771",
    archivePrefix = "arXiv",
    primaryClass = "astro-ph.CO",
    reportNumber = "LIGO-DCC-P1800158-v12",
    doi = "10.1103/PhysRevLett.121.231103",
    journal = "Phys. Rev. Lett.",
    volume = "121",
    number = "23",
    pages = "231103",
    year = "2018"
}

@article{LIGOScientific:2005fbz,
    author = "Abbott, B. and others",
    collaboration = "LIGO Scientific",
    title = "{Search for gravitational waves from primordial black hole binary coalescences in the galactic halo}",
    eprint = "gr-qc/0505042",
    archivePrefix = "arXiv",
    reportNumber = "LIGO-P040045-04-Z",
    doi = "10.1103/PhysRevD.72.082002",
    journal = "Phys. Rev. D",
    volume = "72",
    pages = "082002",
    year = "2005"
}

@article{LIGOScientific:2026XXX,
    author = "Abac, A. G. and others",
    collaboration = "LIGO Scientific, VIRGO, KAGRA",
    title = "{Searches for Binary Mergers with Sub-solar Mass Components in Data from the First Part of LIGO--Virgo--KAGRA's Fourth Observing Run}",
    eprint = "2605.05444",
    journal = "arXiv",
    archivePrefix = "arXiv",
    primaryClass = "astro-ph.HE",
    month = "5",
    year = "2026"
}

@article{Hanna:2022zpk,
    author = "Hanna, Chad and others",
    title = "{Binary tree approach to template placement for searches for gravitational waves from compact binary mergers}",
    eprint = "2209.11298",
    archivePrefix = "arXiv",
    primaryClass = "gr-qc",
    doi = "10.1103/PhysRevD.108.042003",
    journal = "Phys. Rev. D",
    volume = "108",
    number = "4",
    pages = "042003",
    year = "2023"
}

@article{Joshi:2025zdu,
    author = "Joshi, Prathamesh and others",
    title = "{How Many Times Should We Matched Filter Gravitational Wave Data? A Comparison of GstLAL's Online and Offline Performance}",
    eprint = "2505.23959",
    journal = {arXiv},
    archivePrefix = "arXiv",
    primaryClass = "gr-qc",
    month = "5",
    year = "2025"
}

@article{Joshi:2025nty,
    author = "Joshi, Prathamesh and others",
    title = "{New Methods for Offline GstLAL Analyses}",
    eprint = "2506.06497",
    archivePrefix = "arXiv",
    journal = {arXiv},
    primaryClass = "gr-qc",
    month = "6",
    year = "2025"
}

@article{Aubin:2020goo,
    author = "Aubin, F. and others",
    title = "{The MBTA pipeline for detecting compact binary coalescences in the third LIGO{\textendash}Virgo observing run}",
    eprint = "2012.11512",
    archivePrefix = "arXiv",
    primaryClass = "gr-qc",
    doi = "10.1088/1361-6382/abe913",
    journal = "Class. Quant. Grav.",
    volume = "38",
    number = "9",
    pages = "095004",
    year = "2021"
}

@article{Allene:2025saz,
    author = "All{\'e}n{\'e}, Christopher and others",
    title = "{The MBTA pipeline for detecting compact binary coalescences in the fourth LIGO-Virgo-KAGRA observing run}",
    eprint = "2501.04598",
    archivePrefix = "arXiv",
    primaryClass = "gr-qc",
    doi = "10.1088/1361-6382/add234",
    journal = "Class. Quant. Grav.",
    volume = "42",
    number = "10",
    pages = "105009",
    year = "2025"
}

@article{Davies:2020tsx,
    author = "Davies, Gareth S. and Dent, Thomas and T{\'a}pai, M{\'a}rton and Harry, Ian and McIsaac, Connor and Nitz, Alexander H.",
    title = "{Extending the PyCBC search for gravitational waves from compact binary mergers to a global network}",
    eprint = "2002.08291",
    archivePrefix = "arXiv",
    primaryClass = "astro-ph.HE",
    doi = "10.1103/PhysRevD.102.022004",
    journal = "Phys. Rev. D",
    volume = "102",
    number = "2",
    pages = "022004",
    year = "2020"
}

@article{LIGOScientific:2025yae,
    author = "Abac, A. G. and others",
    collaboration = "LIGO Scientific, VIRGO, KAGRA",
    title = "{GWTC-4.0: Methods for Identifying and Characterizing Gravitational-wave Transients}",
    eprint = "2508.18081",
    archivePrefix = "arXiv",
    primaryClass = "gr-qc",
    reportNumber = "LIGO-P2400300",
    doi = "10.3847/2041-8213/ae447b",
    journal = "Astrophys. J. Lett.",
    volume = "1004",
    number = "2",
    pages = "L21",
    year = "2026"
}

@article{Khan:2015jqa,
    author = {Khan, Sebastian and Husa, Sascha and Hannam, Mark and Ohme, Frank and P{\"u}rrer, Michael and Jim{\'e}nez Forteza, Xisco and Boh{\'e}, Alejandro},
    title = "{Frequency-domain gravitational waves from nonprecessing black-hole binaries. II. A phenomenological model for the advanced detector era}",
    eprint = "1508.07253",
    archivePrefix = "arXiv",
    primaryClass = "gr-qc",
    doi = "10.1103/PhysRevD.93.044007",
    journal = "Phys. Rev. D",
    volume = "93",
    number = "4",
    pages = "044007",
    year = "2016"
}

@article{Husa:2015iqa,
    author = {Husa, Sascha and Khan, Sebastian and Hannam, Mark and P{\"u}rrer, Michael and Ohme, Frank and Jim{\'e}nez Forteza, Xisco and Boh{\'e}, Alejandro},
    title = "{Frequency-domain gravitational waves from nonprecessing black-hole binaries. I. New numerical waveforms and anatomy of the signal}",
    eprint = "1508.07250",
    archivePrefix = "arXiv",
    primaryClass = "gr-qc",
    doi = "10.1103/PhysRevD.93.044006",
    journal = "Phys. Rev. D",
    volume = "93",
    number = "4",
    pages = "044006",
    year = "2016"
}

@article{Buonanno:2009zt,
    author = "Buonanno, Alessandra and Iyer, Bala and Ochsner, Evan and Pan, Yi and Sathyaprakash, B. S.",
    title = "{Comparison of post-Newtonian templates for compact binary inspiral signals in gravitational-wave detectors}",
    eprint = "0907.0700",
    archivePrefix = "arXiv",
    primaryClass = "gr-qc",
    doi = "10.1103/PhysRevD.80.084043",
    journal = "Phys. Rev. D",
    volume = "80",
    pages = "084043",
    year = "2009"
}

@article{Boyle:2009dg,
    author = "Boyle, Michael and Brown, Duncan A. and Pekowsky, Larne",
    editor = "Sutton, Patrick and Shoemaker, Deirdre",
    title = "{Comparison of high-accuracy numerical simulations of black-hole binaries with stationary phase post-Newtonian template waveforms for Initial and Advanced LIGO}",
    eprint = "0901.1628",
    archivePrefix = "arXiv",
    primaryClass = "gr-qc",
    doi = "10.1088/0264-9381/26/11/114006",
    journal = "Class. Quant. Grav.",
    volume = "26",
    pages = "114006",
    year = "2009"
}

@Inbook{Isoyama2020,
author="Isoyama, Soichiro
and Sturani, Riccardo
and Nakano, Hiroyuki",
editor="Bambi, Cosimo
and Katsanevas, Stavros
and Kokkotas, Konstantinos D.",
title="{Post-Newtonian Templates for Gravitational Waves from Compact Binary Inspirals}",
bookTitle="Handbook of Gravitational Wave Astronomy",
year="2020",
publisher="Springer Singapore",
address="Singapore",
pages="1--49",
isbn="978-981-15-4702-7",
doi="10.1007/978-981-15-4702-7_31-1",
url="https://doi.org/10.1007/978-981-15-4702-7_31-1"
}

@article{PhysRevD.101.124040,
  title = {Frequency-domain reduced-order model of aligned-spin effective-one-body waveforms with higher-order modes},
  author = {Cotesta, Roberto and Marsat, Sylvain and P\"urrer, Michael},
  journal = {Phys. Rev. D},
  volume = {101},
  issue = {12},
  pages = {124040},
  numpages = {17},
  year = {2020},
  month = {Jun},
  publisher = {American Physical Society},
  doi = {10.1103/PhysRevD.101.124040},
  url = {https://link.aps.org/doi/10.1103/PhysRevD.101.124040}
}

@article{PhysRevLett.57.2607,
  title = "{Replica Monte Carlo Simulation of Spin-Glasses}",
  author = {Swendsen, Robert H. and Wang, Jian-Sheng},
  journal = {Phys. Rev. Lett.},
  volume = {57},
  issue = {21},
  pages = {2607--2609},
  numpages = {0},
  year = {1986},
  month = {Nov},
  publisher = {American Physical Society},
  doi = {10.1103/PhysRevLett.57.2607},
  url = {https://link.aps.org/doi/10.1103/PhysRevLett.57.2607}
}

@article{LIGOScientific:2025snk,
    author = "Abac, A. G. and others",
    collaboration = "LIGO Scientific, Virgo, KAGRA",
    title = "{Open Data from LIGO, Virgo, and KAGRA through the First Part of the Fourth Observing Run}",
    eprint = "2508.18079",
    archivePrefix = "arXiv",
    primaryClass = "gr-qc",
    reportNumber = "LIGO-P2500167",
    doi = "10.3847/1538-4357/ae211e",
    journal = "Astrophys. J.",
    volume = "1004",
    number = "2",
    pages = "232",
    year = "2026"
}

@article{Niu:2025nha,
    author = "Niu, Wanting and others",
    title = "{GW231109{\_}235456: A Sub-threshold Binary Neutron Star Merger in the LIGO-Virgo-KAGRA O4a Observing Run?}",
    eprint = "2509.09741",
    archivePrefix = "arXiv",
    journal = {arXiv},
    primaryClass = "astro-ph.HE",
    reportNumber = "LIGO-P2500544",
    month = "9",
    year = "2025"
}

@article{Kacanja:2026byy,
    author = {Kacanja, Keisi and Soni, Kanchan and Aky{\"u}z, Aleyna and Nitz, Alexander H.},
    title = "{Search for subsolar mass binaries in the first part of LIGO{\textquoteright}s fourth observing run}",
    eprint = "2602.12115",
    archivePrefix = "arXiv",
    primaryClass = "astro-ph.HE",
    doi = "10.1103/ztpt-pwl5",
    journal = "Phys. Rev. D",
    volume = "114",
    number = "4",
    pages = "043030",
    year = "2026"
}

@article{Prunier:2023uoo,
    author = "Prunier, Marine and Morr{\'a}s, Gonzalo and Siles, Jos{\'e} Francisco Nu{\~n}o and Clesse, Sebastien and Garc{\'\i}a-Bellido, Juan and Ruiz Morales, Ester",
    title = "{Analysis of the subsolar-mass black hole candidate SSM200308 from the second part of the third observing run of Advanced LIGO-Virgo}",
    eprint = "2311.16085",
    archivePrefix = "arXiv",
    primaryClass = "gr-qc",
    doi = "10.1016/j.dark.2024.101582",
    journal = "Phys. Dark Univ.",
    volume = "46",
    pages = "101582",
    year = "2024"
}

@article{LIGOScientific:2026wfs,
    collaboration = "LIGO Scientific, VIRGO, KAGRA",
    title = "{GWTC-5.0: Observations from the Second Part of the Fourth LIGO-Virgo-KAGRA Observing Run and Updates to the Gravitational-Wave Transient Catalog}",
    eprint = "2605.27225",
    archivePrefix = "arXiv",
    journal = "arXiv",
    primaryClass = "gr-qc",
    reportNumber = "LIGO-P2600152",
    month = "5",
    year = "2026"
}

@article{LIGOScientific:2026sit,
    author = "Abac, None and others",
    collaboration = "LIGO Scientific, VIRGO, KAGRA",
    title = "{GWTC-5.0: An Introduction to Version 5.0 of the Gravitational-Wave Transient Catalog}",
    eprint = "2605.27223",
    archivePrefix = "arXiv",
    journal = "arXiv",
    primaryClass = "gr-qc",
    reportNumber = "LIGO-P2500701",
    month = "5",
    year = "2026"
}

@article{LIGOScientific:2017vwq,
    author = "Abbott, B. P. and others",
    collaboration = "LIGO Scientific, Virgo",
    title = "{GW170817: Observation of Gravitational Waves from a Binary Neutron Star Inspiral}",
    eprint = "1710.05832",
    archivePrefix = "arXiv",
    primaryClass = "gr-qc",
    reportNumber = "LIGO-P170817",
    doi = "10.1103/PhysRevLett.119.161101",
    journal = "Phys. Rev. Lett.",
    volume = "119",
    number = "16",
    pages = "161101",
    year = "2017"
}

@article{LIGOScientific:2020aai,
    author = "Abbott, B. P. and others",
    collaboration = "LIGO Scientific, Virgo",
    title = "{GW190425: Observation of a Compact Binary Coalescence with Total Mass $\sim 3.4 M_{\odot}$}",
    eprint = "2001.01761",
    archivePrefix = "arXiv",
    primaryClass = "astro-ph.HE",
    reportNumber = "LIGO-P190425",
    doi = "10.3847/2041-8213/ab75f5",
    journal = "Astrophys. J. Lett.",
    volume = "892",
    number = "1",
    pages = "L3",
    year = "2020"
}

@article{LIGOScientific:2021qlt,
    author = "Abbott, R. and others",
    collaboration = "LIGO Scientific, KAGRA, VIRGO",
    title = "{Observation of Gravitational Waves from Two Neutron Star{\textendash}Black Hole Coalescences}",
    eprint = "2106.15163",
    archivePrefix = "arXiv",
    primaryClass = "astro-ph.HE",
    reportNumber = "LIGO Document P2000357",
    doi = "10.3847/2041-8213/ac082e",
    journal = "Astrophys. J. Lett.",
    volume = "915",
    number = "1",
    pages = "L5",
    year = "2021"
}

@article{LIGOScientific:2024elc,
    author = "Abac, A. G. and others",
    collaboration = "LIGO Scientific, KAGRA, VIRGO",
    title = "{Observation of Gravitational Waves from the Coalescence of a 2.5{\textendash}4.5 M$_{\odot}$ Compact Object and a Neutron Star}",
    eprint = "2404.04248",
    archivePrefix = "arXiv",
    primaryClass = "astro-ph.HE",
    reportNumber = "LIGO-P2300352",
    doi = "10.3847/2041-8213/ad5beb",
    journal = "Astrophys. J. Lett.",
    volume = "970",
    number = "2",
    pages = "L34",
    year = "2024"
}

@article{LIGOScientific:2020zkf,
    author = "Abbott, R. and others",
    collaboration = "LIGO Scientific, Virgo",
    title = "{GW190814: Gravitational Waves from the Coalescence of a 23 Solar Mass Black Hole with a 2.6 Solar Mass Compact Object}",
    eprint = "2006.12611",
    archivePrefix = "arXiv",
    primaryClass = "astro-ph.HE",
    reportNumber = "LIGO-P190814",
    doi = "10.3847/2041-8213/ab960f",
    journal = "Astrophys. J. Lett.",
    volume = "896",
    number = "2",
    pages = "L44",
    year = "2020"
}

@article{Cornish:2020dwh,
    author = "Cornish, Neil J. and Littenberg, Tyson B. and B{\'e}csy, Bence and Chatziioannou, Katerina and Clark, James A. and Ghonge, Sudarshan and Millhouse, Margaret",
    title = "{BayesWave analysis pipeline in the era of gravitational wave observations}",
    eprint = "2011.09494",
    archivePrefix = "arXiv",
    primaryClass = "gr-qc",
    doi = "10.1103/PhysRevD.103.044006",
    journal = "Phys. Rev. D",
    volume = "103",
    number = "4",
    pages = "044006",
    year = "2021"
}

@article{Cornish_2015,
   title={Bayeswave: Bayesian inference for gravitational wave bursts and instrument glitches},
   volume={32},
   ISSN={1361-6382},
   url={http://dx.doi.org/10.1088/0264-9381/32/13/135012},
   DOI={10.1088/0264-9381/32/13/135012},
   number={13},
   journal={Classical and Quantum Gravity},
   publisher={IOP Publishing},
   author={Cornish, Neil J and Littenberg, Tyson B},
   year={2015},
   month=June, pages={135012} }

@article{PhysRevD.91.084034,
  title = {Bayesian inference for spectral estimation of gravitational wave detector noise},
  author = {Littenberg, Tyson B. and Cornish, Neil J.},
  journal = {Phys. Rev. D},
  volume = {91},
  issue = {8},
  pages = {084034},
  numpages = {13},
  year = {2015},
  month = {Apr},
  publisher = {American Physical Society},
  doi = {10.1103/PhysRevD.91.084034},
  url = {https://link.aps.org/doi/10.1103/PhysRevD.91.084034}
}

@article{PhysRevD.109.064040,
  title = {Bayesian power spectral estimation of gravitational wave detector noise revisited},
  author = {Gupta, Toral and Cornish, Neil J.},
  journal = {Phys. Rev. D},
  volume = {109},
  issue = {6},
  pages = {064040},
  numpages = {11},
  year = {2024},
  month = {Mar},
  publisher = {American Physical Society},
  doi = {10.1103/PhysRevD.109.064040},
  url = {https://link.aps.org/doi/10.1103/PhysRevD.109.064040}
}

@article{müller2025minimumneutronstarmass,
    author = {M{\"u}ller, Bernhard and Heger, Alexander and Powell, Jade},
    title = "{Minimum Neutron Star Mass in Neutrino-Driven Supernova Explosions}",
    eprint = "2407.08407",
    archivePrefix = "arXiv",
    primaryClass = "astro-ph.HE",
    doi = "10.1103/PhysRevLett.134.071403",
    journal = "Phys. Rev. Lett.",
    volume = "134",
    number = "7",
    pages = "071403",
    year = "2025"
}

@misc{chen2025gravitationalinstabilityfragmentationcollapsar,
      title="{Gravitational Instability and Fragmentation in Collapsar Disks Supports the Formation of Sub-Solar Neutron Stars}", 
      author={Yi-Xian Chen and Brian D. Metzger},
      year={2025},
      eprint={2508.17183},
      archivePrefix={arXiv},
      primaryClass={astro-ph.HE},
      url={https://arxiv.org/abs/2508.17183}, 
}

@article{Wolfe:2023yuu,
    author = "Wolfe, Noah E. and Vitale, Salvatore and Talbot, Colm",
    title = "{Too small to fail: characterizing sub-solar mass black hole mergers with gravitational waves}",
    eprint = "2305.19907",
    archivePrefix = "arXiv",
    primaryClass = "astro-ph.HE",
    doi = "10.1088/1475-7516/2023/11/039",
    journal = "JCAP",
    volume = "11",
    pages = "039",
    year = "2023"
}

@article{Newell:2026cma,
    author = "Newell, Murdoc and Boudon, Alexis and Qi, Hong",
    title = "{Multibanded Reduced Order Quadrature Techniques for Gravitational Wave Inference}",
    eprint = "2601.09819",
    archivePrefix = "arXiv",
    journal = "arXiv",
    primaryClass = "gr-qc",
    month = "1",
    year = "2026"
}

\end{document}